\documentclass[a4paper]{article}
\pdfoutput=1

\usepackage[english]{babel}
\usepackage[utf8]{inputenc}
\usepackage[T1]{fontenc}
\usepackage{amsmath}
\usepackage{amssymb}
\usepackage{graphicx}
\usepackage[colorinlistoftodos]{todonotes}
\usepackage[hyphens]{url}
\usepackage[colorlinks,urlcolor=blue]{hyperref}
\usepackage{listings}
\usepackage{color}
\usepackage{graphicx}
\usepackage{caption}
\usepackage{subcaption}
\usepackage{paralist}
\usepackage{csquotes}
\usepackage{authblk}

\title{Benchmarking Cognitive Abilities of the Brain with Computer Games}

\author[1,*]{Norbert B\'atfai}
\author[2]{D\'avid Papp} 
\author[1]{Ren\'at\'o Besenczi}
\author[1]{Gerg{\H o} Bogacsovics}
\author[1]{D\'avid Veres}

\affil[1]{Department of Information Technology, University of Debrecen, Hungary}
\affil[2]{Department of Psychology, University of Debrecen, Hungary}
\affil[*]{Corresponding author: Norbert B\'atfai, batfai.norbert@inf.unideb.hu}

\begin{document}
\maketitle
\begin{abstract}
Most of the players have experienced the feeling of temporarily losing their character in a given gameplay situation when they cannot control the character, simply because they temporarily cannot see it. The main reasons for this feeling may be due to the interplay of the following factors: (1) the visual complexity of the game is unexpectedly increased compared with the previous time period as more and more game objects and effects 
are rendered on the display; (2) and/or the game is lagging; (3) and finally, it is also possible that the players have no sufficient experience with controlling the character. This paper focuses on the first reason. We have developed a benchmark program 
which allows its user to experience the feeling of losing character. While the user can control the character well the benchmark program will increase the visual complexity of the display. Otherwise, if the user lost the character then the program will decrease the complexity until the user will find the character again, and so on. The complexity is measured based on the number of changed pixels between two consecutive display images. Our measurements show that the average of bit per second values of losing and finding pairs describes the user well. 
The final goal of this research is to further develop our benchmark to a standard psychological test.
\end{abstract}


\section{Introduction}
\begin{figure*}
  \centering
    \includegraphics[width=\textwidth]{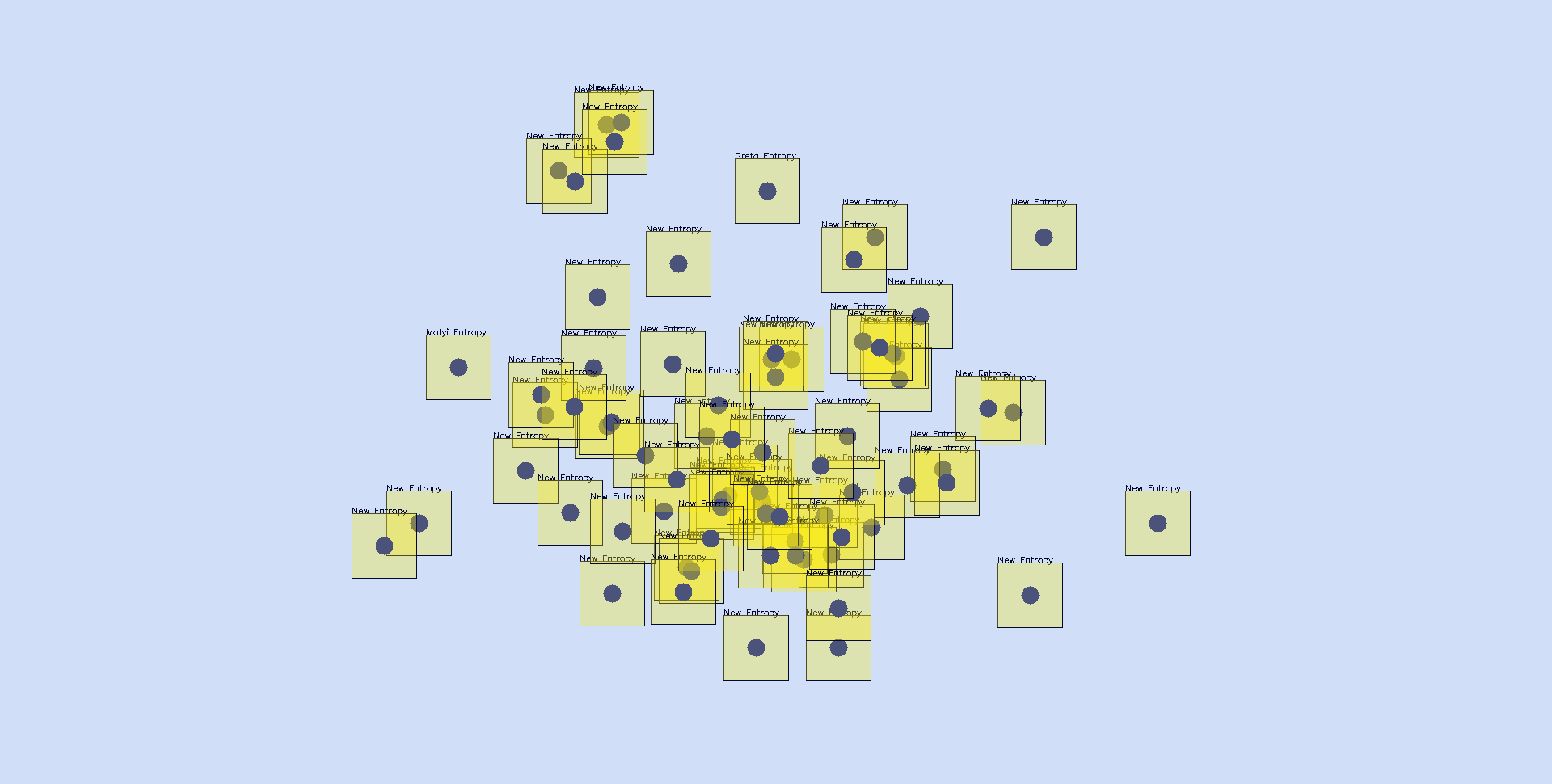}
  \caption{A screenshot of BrainB Test Series 6 in action. Can you find the box labelled by the name Samu Entropy in this picture?}
  \label{brainbs6screen}
\end{figure*}

Losing the control of the character in a given gameplay situation is a very common feeling that is well known among gamers. In this situation, players cannot control their character, simply because they temporarily cannot see it due to the visual complexity of the display is unexpectedly increased and/or the game is lagging and, finally, it is also possible that the players have no sufficient experience to control the character. 
In this paper, we introduce our benchmark computer program called \textit{BrainB Test Series 6} that can abstract this feeling. In this test, game objects are symbolized by boxes as it can be seen in Fig \ref{brainbs6screen}. All boxes move according to random walks. There is a distinguished box labelled by the name \textit{Samu Entropy}. It represents the character controlled by the player. 
The benchmark test lasts for 10 minutes. 
During the test, the user must continuously hold and drag the mouse button on the center of Samu Entropy. 
If the user succeeds in this task then the benchmark program will increase the visual complexity of the display. It will draw more and more overlapping boxes which will move faster and faster. Otherwise, if the mouse pointer cannot follow the center of Samu Entropy then the visual complexity will be decreased. The test will delete more and more boxes and the remaining boxes move slower 
and slower until the user finds Samu Entropy again, i.e., clicks on Samu Entropy.

The BrainB Series 1 to 4 were developed in the family setting of the first author\footnote{For example, see \url{https://www.twitch.tv/videos/139186614}}. 
Then, in our university environment, 
we had already done a preliminary study \cite{brainbs5} on a previous (BrainB Series 5) version of our benchmark. Some of its measurements were streamed live on Twitch\footnote{For example, see \url{https://www.twitch.tv/videos/206478952}}. 
The main research goal of this study is to show that players lose the character on a higher complexity level of the display and they find it on a relatively lower complexity level.

The organization of the paper is the following. In the next section, we give a brief overview of the psychological and informatics background and the phenomenon of losing the character is illustrated. The second section presents the algorithm and the operation of our benchmark program including the presentation of the first measurements followed by systematic measurements. 
Finally, we conclude our paper and show some future plans.

\subsection{Psychological Background}
The cognitive ability of attention is a significant factor in everyday life, either it comes from work, hobby or the daily activities, 
as it affects the performance of all the previously mentioned things. 
The alertness, or in other words, the long upheld attention, in technical terms is called vigilance. 
The research of vigilance is an important topic in Psychology from 1970 to the present day. 
The first method used was the Mackworth Clock \cite{Mackworth}, in which the participants had to pay attention to a clock that had a second hand which sometimes sprang twice, and then the participants had to signal as soon as possible. 
For measuring attention and concentration, there is another method, 
the Toulouse-Piéron test \cite{ToulousePieron}, 
in which participants have to follow a given scheme to separate right and wrong signs. To measure vigilance we must take into consideration the hit ratio and the number of false alarms. In almost all of our activities there are also interfering stimuli that affects our performance as well. These other factors vary by quantity and quality, and some can be stimulating, 
while some detain us 
from the optimal performance. The Yerkes-Dodson law \cite{YerkesDodsonLaw} says that for achieving the best performance 
there is an optimal arousal level, which level is higher in simpler tasks, and lower in complex activities. 
It can be represented by a inverted U-shaped curve. 
We must not forget that as in some other things, in the attentional system there are also personal differences that 
should be taken into consideration while researching the subject \cite{AltPszichobook}. 

Other objects in the environment can affect how we perceive the one object that is interesting for us. 
In 1940, 
Witkin et al.\ did a research on perception \cite{WitkinEtAl}, and from this work, they created a theory about two different cognitive styles, which they called field dependent, and field independent. A field dependent person perception is mainly affected by the field, the environment of the observed object. On the contrary, a field independent person does not affected by the field created by the observed object’s environment. This phenomenon was investigated by a task, in which the participants had to determine whether a straight rod, 
in different planes is vertical or not.
Moreover, there is another typical method used in this topic, that is the Tilting room, Tilting chair test, 
in which the participant is sitting in a tiltable chair that he or she needs to controll in order to get him/herself into vertical position despite the tilting room. 
Later, Witkin and Goodenough reinvestigated the topic, 
and they came to a conclusion that the two styles 
are two ends of the spectrum, 
however, some people are fixed with one of the cognitive styles, 
while others can adapt to the style they use depending on the situation \cite{WitkinGoodenough}. 

Speaking of attention, it’s important to talk about the main processing system, i.e., the brain. The operation of the brain is frequently compared to the mechanism of a personal computer 
by many researchers. 
Carl Sagan based his theory on the binary coding, 
so he used the information content in binary. 
When we are watching something, the picture seen that our brain maps, is made of plenty of information. 
Sagan wanted to calculate the 
information processing 
speed of the brain, to do so, he based his calculation on the example of looking at the moon, and from this example he drew the consequence, that the brain can process about 5000 bit/sec at its peak performance \cite{Sagan}. In a modern project, called Building 8, the main thought is to make the brain into a computer. Based on this project, the
information processing speed of the brain 
is about a terrabyte/sec, which far exceeds the speed estimated by Sagan \cite{Nieva}.  

\subsubsection{Practicing filling out tests}

Filling out tests and experiments are common tools in the science of psychology. Countless methods were 
created to date, but these methods are not just used, because researchers improve them, as well as, try to test them in a wider range. However, we need to consider certain factors in each experiment 
and test
that how they affect the method’s usability and the final results as well. Among these factors, there is one, when the participant obtains knowledge about what is expected from her/him, 
or which answer are considered the ’best’. This way the participant will accomodate to this information, because he/she, as everyone else, wants to portray herself/himself in the best manner possible and to be the ’best’ in performance. In multiple choice questions, there are some tricks, that are well known in the common knowledge, which we all use, when we don’t know the right answer for sure. A somewhat similar tool is the experience or 
routine with filling out tests, which can help to choose the adequate strategy for solving the situation, this is called test-wisdom. To achieve that, one must discover the logic behind the method, or practise it many times. But the test-wisdom often cause inconvenience for  
test developers, because they have to keep in mind a bonus factor, which is totally diverge from the basic variables they meant to manipulate, and vary in each individual \cite{PszichoMeresbook}.

The effect of being experienced in filling in tests was studied in a research, in which an aptitude test called GRE (Grand Record Examination) was used. Practise samples were sent to random participants 5 weeks before the examination. Those who got these samples also 
receieved advices for completing. In conclusion, the group with prior knowledge and practise earned significantly better results in the examination. Furthermore, there were also a notable growth in points, when the participants received an only 4 hours educational practise before the examination. It’s important to note that this difference and growth was present only in the logical reasoning part of the exam, and not in the mathematical and verbal parts 
\cite{PowersSwinton}. This data was reexamined later, because researchers wanted to know, if there is any difference when the existing groups would be split into subgroups by the different attributes of the participants. As a conclusion, there was no significant difference between the subgroups, but there was a notable difference in the group in which the members' primary language was not English, they scored lesser points than the others \cite{Powers86}.

Repeatedly performing the same experiment or test with the same participants could affect the results. Previously, as we specified, repeatedly using the same method could cause the lowering of its validity, and the results could be distorted. Participants can learn and adapt to certain methods, even if its just means a small percentage of difference. The current test takes 10 minutes to complete, in this 10 minutes the participant’s full attention and concentration is needed.  We should keep in mind, that the negative effects of fatigue could balance the positive effects of practise, in a direct way of repeated examinations. So this two factors should be considered in the evaulation, and while drawing consequence.

It is therefore proposed to perform our benchmark test in a competitive way trying to beat friends, family members, colleagues or ourselves.

\subsection{Informatics Background}
Since computer games have a relatively short history and their effects on cognitive skills have just been started to be researched recently, there are plenty of questions to be answered. In \cite{HOMER201850}, authors reported an increase in executive functions in school students after playing computer games. Moisala et al. in \cite{MOISALA2017204} shows that enhancements in speed and performance accuracy 
of working memory tasks is related to daily gaming activity. In \cite{bediou2018meta}, authors present an analysis of the impact of action video games on cognitive skills.

Using computer games to measure cognitive abilities has a short history, but a promising future. Most research try to measure the presence or severity of a certain cognitive disease such as dementia or Alzheimer's disease.
In \cite{anguera2013video}, authors show how a long-term use of video games can reduce multitasking costs in older adults. Geyer et al. in \cite{GEYER2015260} show that the change of the score of an online game is in connection with the age-related changes in working memory.

Seldom can we find applications that has been developed for the measurement of cognitive abilities. One such application is reported in \cite{10.1007/978-3-319-27695-3_13} and \cite{pataki2015computer}, it is a framework that has been developed to measure cognitive abilities and its change of elders with computer games. This framework is able to log and analyze scores achieved in various online computer games.

From the viewpoint of information theory and HCI (Human-Computer Interaction), the Hick's law \cite{seow} could be an interesting aspect. This law states that the response time of the brain increases with logarithm of the size of the input. For our purposes, it can be an interesting question: 
how can we apply the Hick's law (or other information theory figure) in our benchmark software?

\subsection{Losing The Character}
We have experienced the feeling of losing the character during playing several games like for example 
League of Legends\footnote{\url{https://na.leagueoflegends.com}}, 
Clash of Clans\footnote{\url{http://supercell.com/en/games/clashofclans/}},
Clash Royale\footnote{\url{http://supercell.com/en/games/clashroyale/}},
Heroes of the Storm\footnote{\url{https://heroesofthestorm.com}}, 
Dota 2\footnote{\url{https://www.dota2.com}}, 
World of Warcraft\footnote{\url{https://worldofwarcraft.com}} 
or Cabal\footnote{\url{http://cabal.playthisgame.com}}.

Now we share our thoughts about the phenomenon of „losing the character”, and give some examples to illustrate it from the game called League of Legends.
Basically, a match starts kind of slowly and quietly: the laners are farming, as well as the junglers in their own territory. Of course smaller fights can occur in the early stages of the game, like a 1v1 in the solo lanes, or a 3v3 in the bottom lane as both of the junglers decides to gank, but these situations are relatively easy to see through. As we head into the mid and late game, teams start fights more often with more people, even with all of them. This is what we call teamfights. These are harder to handle, 
because a lot of things can appear on our screen at the same time: the champions who participate in the fight, optionally minions or jungle monsters, and the visual effects of the spells, summoner spells, and the active or passive abilities of the items. Besides them, we see a lot of things, we still have to make sure that we fulfill our ingame role properly: position well, attack the proper target, or defend our teammates. We have to handle a lot of information at a blink of an eye, so it is completely natural, that sometimes we do not know where to look at or what to do. We can lose our own character, which can end with our death; we can lose the target character, and it can survive; or we can lose the character that we wanted to protect, thus an important member of the team can die.
This can be a short explanation of the phenomenon, which we can also call „losing the focus”.

An example ingame footage can be viewed at \url{https://youtu.be/wdy3KUm1454}, starting at 2:12.

These situations are one of the hardest parts of the game, and it is not easy to handle them well. The easiest way to prepare for them is to play a lot of games, and get experience in them. Also it can help a lot, if we think ahead before a potential teamfight, e.g. which character will be our target, who should we be afraid of, what summoner spells the enemy still has, etc. All of these things can help to execute the fights more properly.

\section{Brain Benchmarking Series 6}

BrainB is a Qt C++ desktop application that uses the OpenCV library. It is developed as an open source project 
that is available on GitHub \cite{brainbsw}. Its source can be built easily on GNU/Linux systems. But the latest (6.0.3) Windows binary release can also be downloaded as a ZIP file from \url{http://smartcity.inf.unideb.hu/~norbi/BrainBSeries6/}.

\begin{figure}
  \centering
     \includegraphics[width=0.75\textwidth]{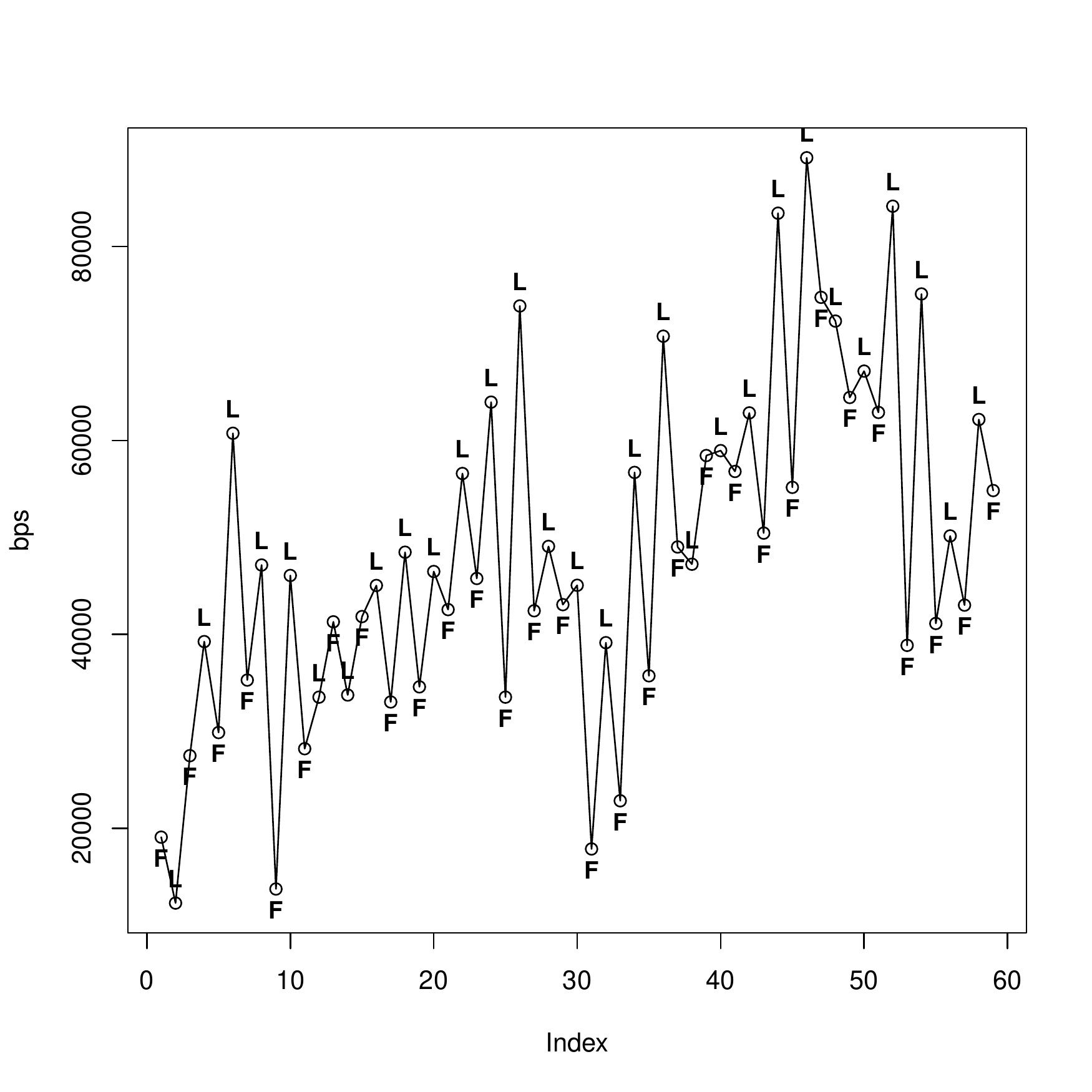}
  \caption{The bps values associated to events of losing and finding. The first element of this sequence is the first element of the \textit{lost2found} (shown in Listing \ref{updadeHeroes} Line \ref{found}) sequence. 
  The second element is the first element of the \textit{found2lost}, and so on. It should be noticed that the 
  losing (labelled by L) and finding (F) events are mixed, see, for example the 13th event on the $x$ axis where the complexity of finding is greather than the complexity of losing in this individual measurement. This test was performed by the first author (46 years old, 
  on a Dell XPS 9333 ultrabook with Windows 10 using the touchpad, 
  resolution 1920x1080, scale 150\%). The final result was 5.92902 Kilobytes. All the logged data can be found at \url{http://smartcity.inf.unideb.hu/~norbi/BrainBSeries6/measurements/NB/}. Fig \ref{brainbs6screen} shows the last screenshot of this experiment.}
   \label{brainbs6winnb1}
\end{figure}

It is important to show the algorithm of BrainB  as precise as possible because the randomness plays a key role in its operation due to boxes doing random walks.
The code snippet shown in Listing \ref{updadeHeroes} is the heart of our benchmark program. 
It is a simplified version of the original source code that can be found in the GitHub repository at  \url{https://github.com/nbatfai/esport-talent-search/blob/master/BrainBWin.cpp#L65}.
This code is executed at every 100 milliseconds that is ten times per second. 
First, as shown in Line \ref{dist}, it computes the distance between the mouse pointer and the center of the box of Samu Entropy and the result is stored in the variable called $dist$ that holds the square of the Euclidean distance. If the distance is larger than $121$ pixels ($11$ is the square root of $121$) and 
if it reoccurs $12$ consecutive times or more in a row (that means at least a time interval of $1.2$ seconds) 
and it is also true that the player was controlling the character well in the previous time slices 
(that is in Line \ref{found} the $state$ is equal to $found$)
then we say that the user has lost the character Samu Entropy and the visual complexity of the display will be saved in Line \ref{save}. The sequence of these losing values and the symmetrical finding values saved in Line \ref{save2} are shown in Fig \ref{brainbs6winnb1}. The complexity is computed in bits per second (bps) units that is based on the number of changed pixels between two consecutive rectangular environments of the character with a given width and height.


\begin{lstlisting}[language={C++},numbers={left},stepnumber={1},numbersep={-1pt},basicstyle={\scriptsize\ttfamily},caption={The algorithm for administration of  losing and finding the character.},label={updadeHeroes},escapechar={|}]
 int dist = ( this->mouse_x - x ) * ( this->mouse_x - x ) |\label{dist}|
  + ( this->mouse_y - y ) * ( this->mouse_y - y );

 if ( dist > 121 ) 
  {
    ++nofLost;
    nofFound = 0;
    if ( nofLost > 12 ) 
      {
        if ( state == found && firstLost ) |\label{found}|
          {
            found2lost.push_back(brainBThread->get_bps()); |\label{save}|
          }
        firstLost = true; |\label{foundlost}|
        state = lost;
        nofLost = 0;
        brainBThread->decComp();
      }
  } 
 else 
  {
    ++nofFound;
    nofLost = 0;
    if ( nofFound > 12 ) 
      {
        if ( state == lost && firstLost ) |\label{lost}|
          {
            lost2found.push_back(brainBThread->get_bps()); |\label{save2}|
          }
        state = found;|\label{lostfound}|
        nofFound = 0;
        brainBThread->incComp();
      }
  }
\end{lstlisting}

The final result printed by the benchmark after it ends in the form \enquote{\textit{U R about 5.92902 Kilobytes}} is the mean of upper bounds for the bps values of the display measured when the variable state changes from found to lost (in Listing \ref{updadeHeroes} from Line \ref{found} to \ref{foundlost}) and vice versa, when the variable state changes from lost to found (in Listing \ref{updadeHeroes} from Lines \ref{lost} to \ref{lostfound}). The simple calculation of this final result is shown in Listing \ref{bps}.

\begin{lstlisting}[language={C++},numbers={left},stepnumber={1},numbersep={-1pt},basicstyle={\scriptsize\ttfamily},caption={The calculation of the final result of the benchmark that is produced in a text file that is saved in the folder where the benchmark was started.},label={bps},escapechar={|}]
  int m1 = mean ( lost2found );
  int m2 = mean ( found2lost );
  
  double res = ( ( ( ( double ) m1
     + ( double ) m2 ) /2.0 ) /8.0 ) /1024.0;
     
  textStream << "U R about " << res << " Kilobytes\n";
\end{lstlisting}

\subsection{First Measurements}

\begin{figure*}
    \centering
    \begin{subfigure}{.45\linewidth}
        \centering
        \includegraphics[scale=0.37]{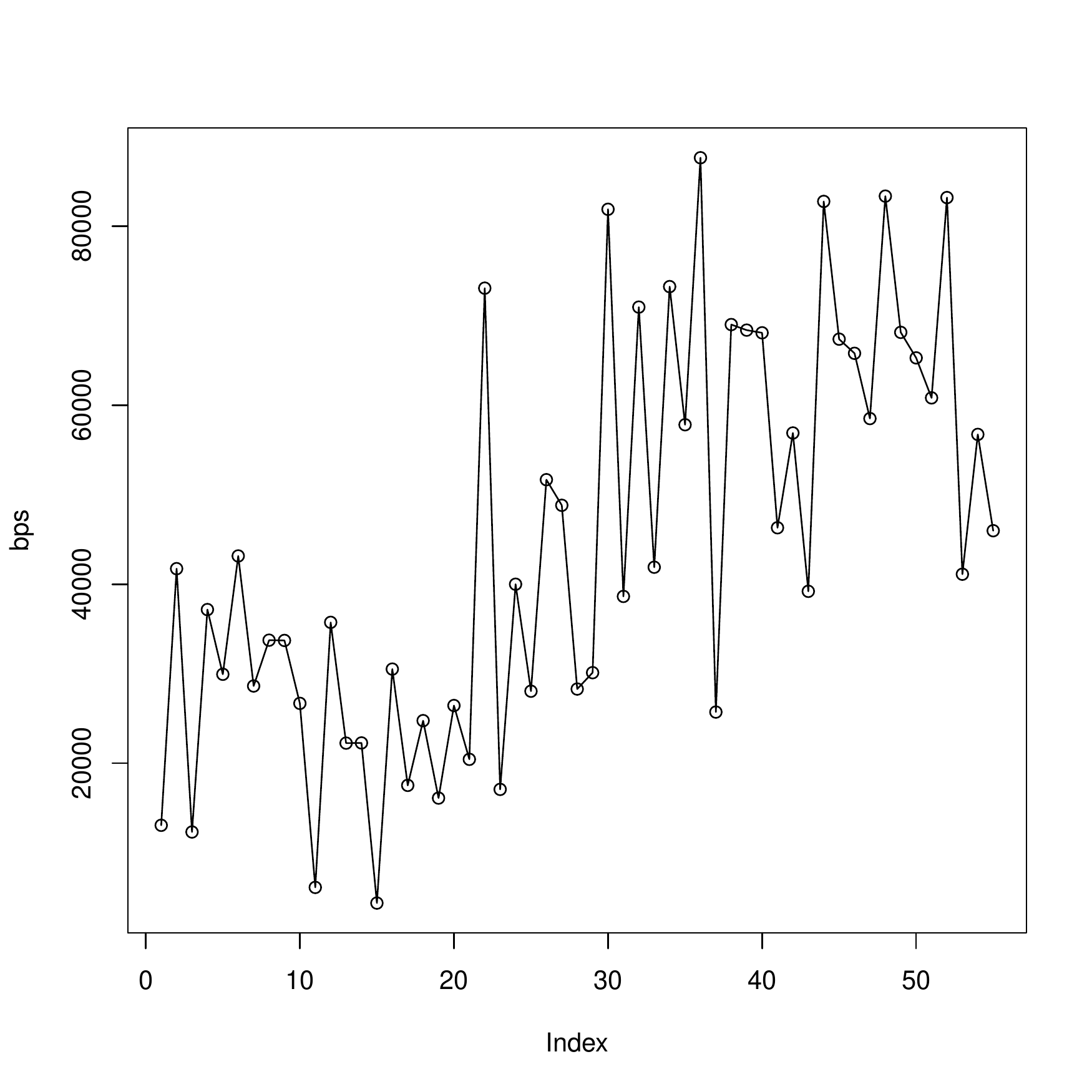}
        \caption{With using the touchpad.}
    \end{subfigure}
        \hskip2em
    \begin{subfigure}{.45\linewidth}
        \centering
        \includegraphics[scale=0.37]{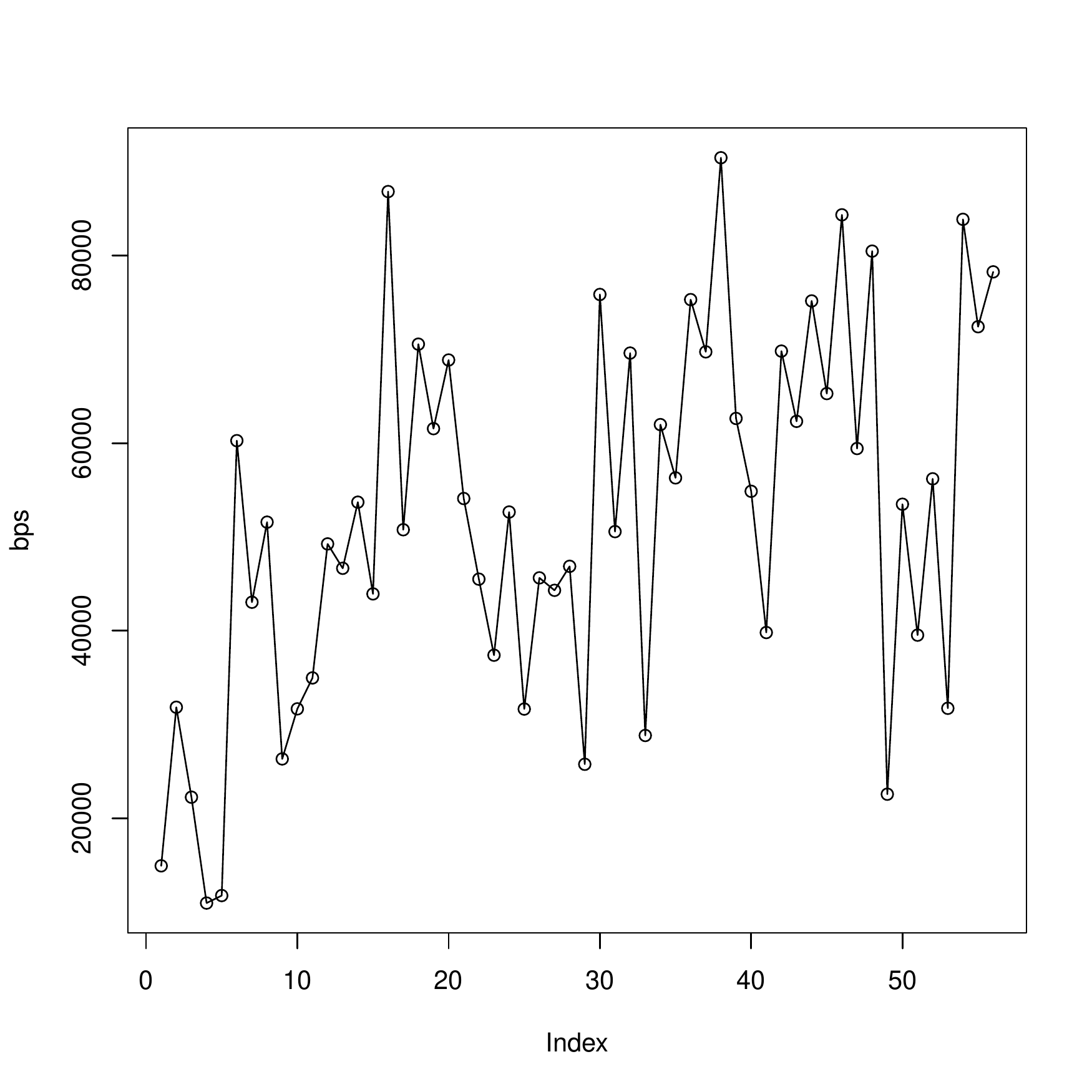}
        \caption{With using a standalone mouse.}
    \end{subfigure}
    \\
    \centering
    \begin{subfigure}{.45\linewidth}
            \centering
        \includegraphics[scale=0.18]{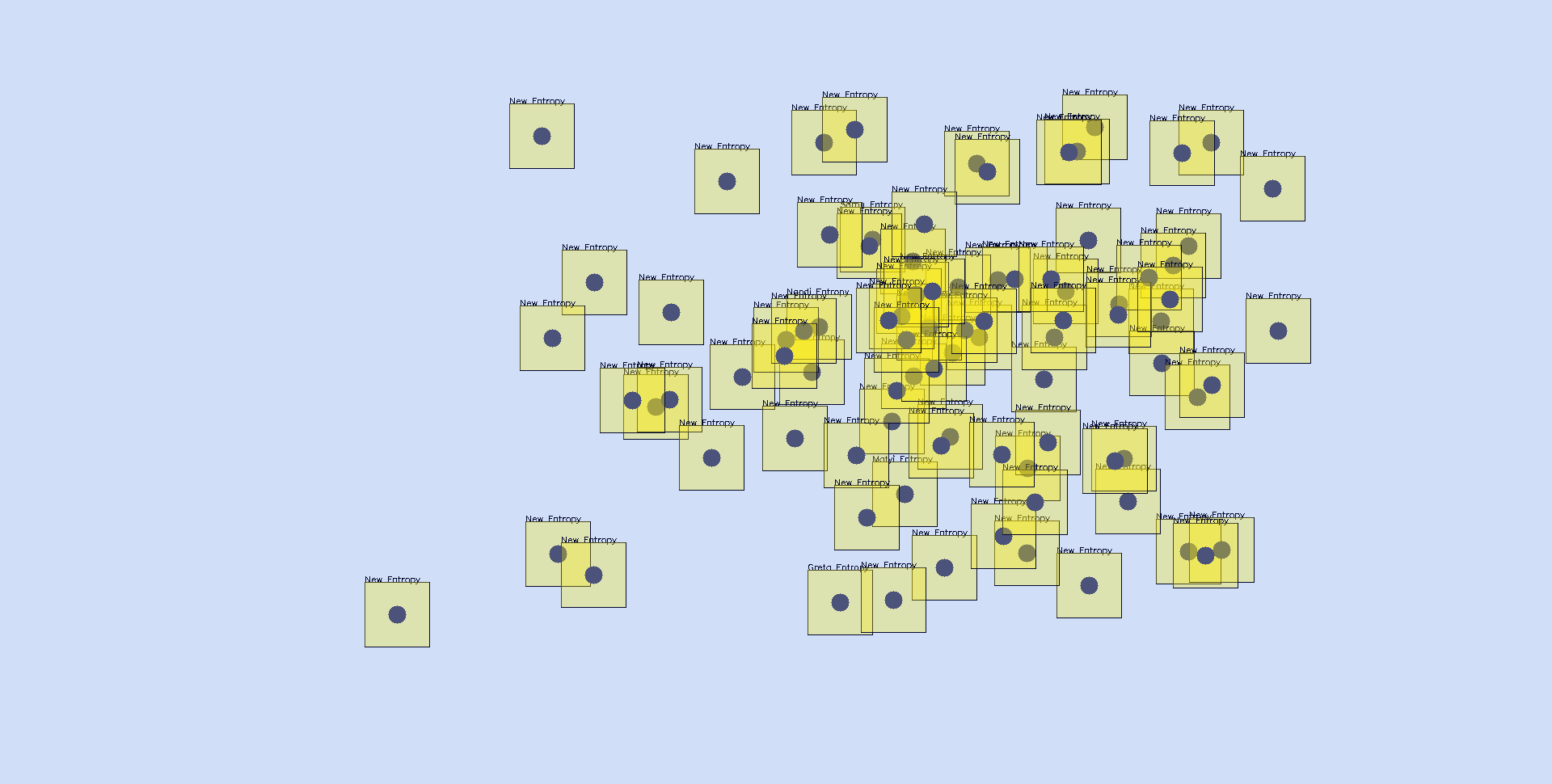}
        \caption{The final result was 5.45563 Kilobytes.}
    \end{subfigure}
    \hskip2em
    \begin{subfigure}{.45\linewidth}
            \centering
        \includegraphics[scale=0.18]{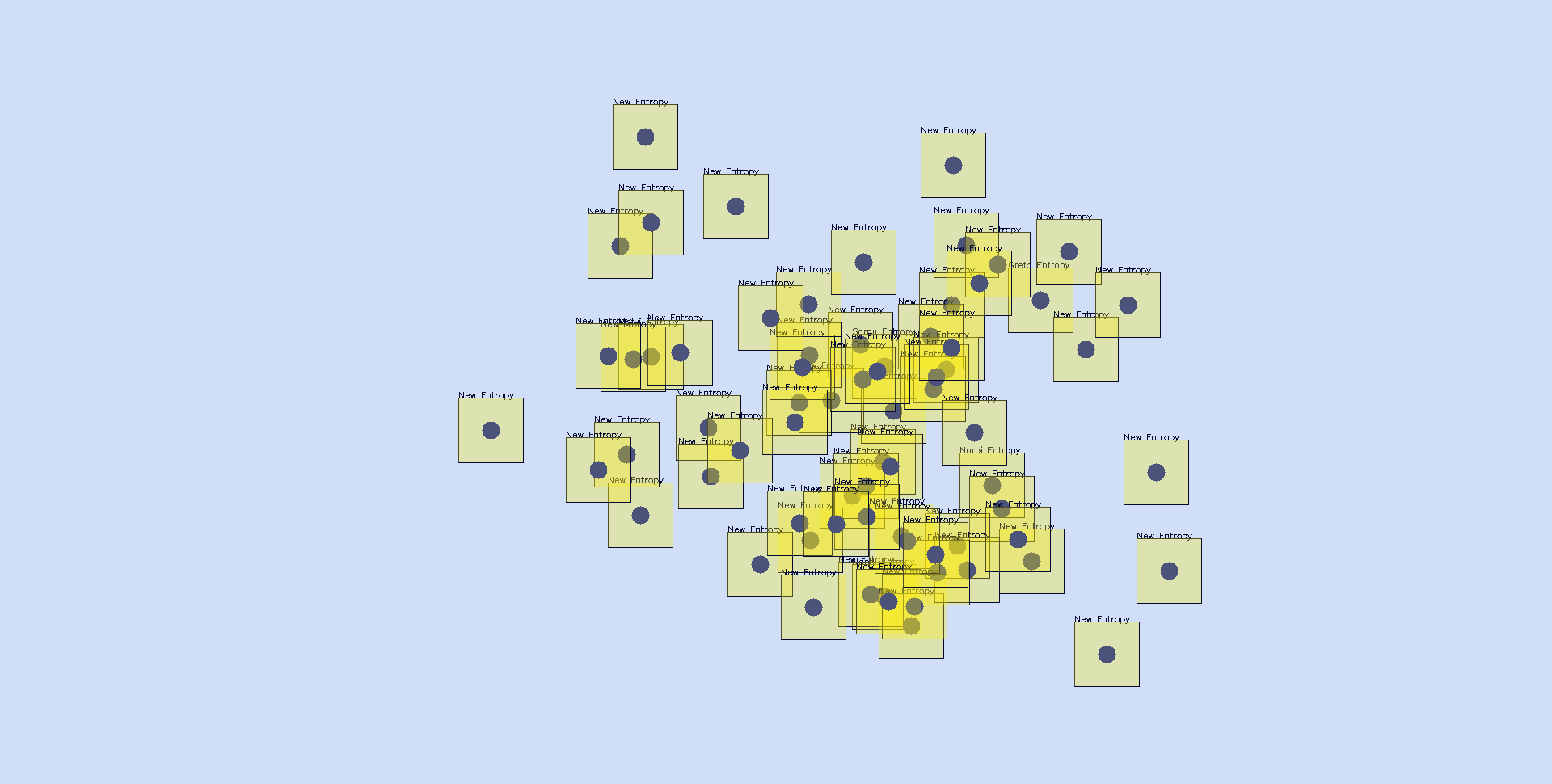}
        \caption{The final result was 6.37927 Kilobytes.}
         \label{brainbs6nblog}
    \end{subfigure}
    \caption{These tests were also performed by the first author on the same environment as in Fig  \ref{brainbs6winnb1}. All the logged data and final screenshots can be found at \url{http://smartcity.inf.unideb.hu/~norbi/BrainBSeries6/measurements/NB/}.}
 \label{brainbs6nb}
\end{figure*}

\begin{figure*}
    \centering
    \begin{subfigure}{.45\linewidth}
        \centering
\includegraphics[scale=0.37]{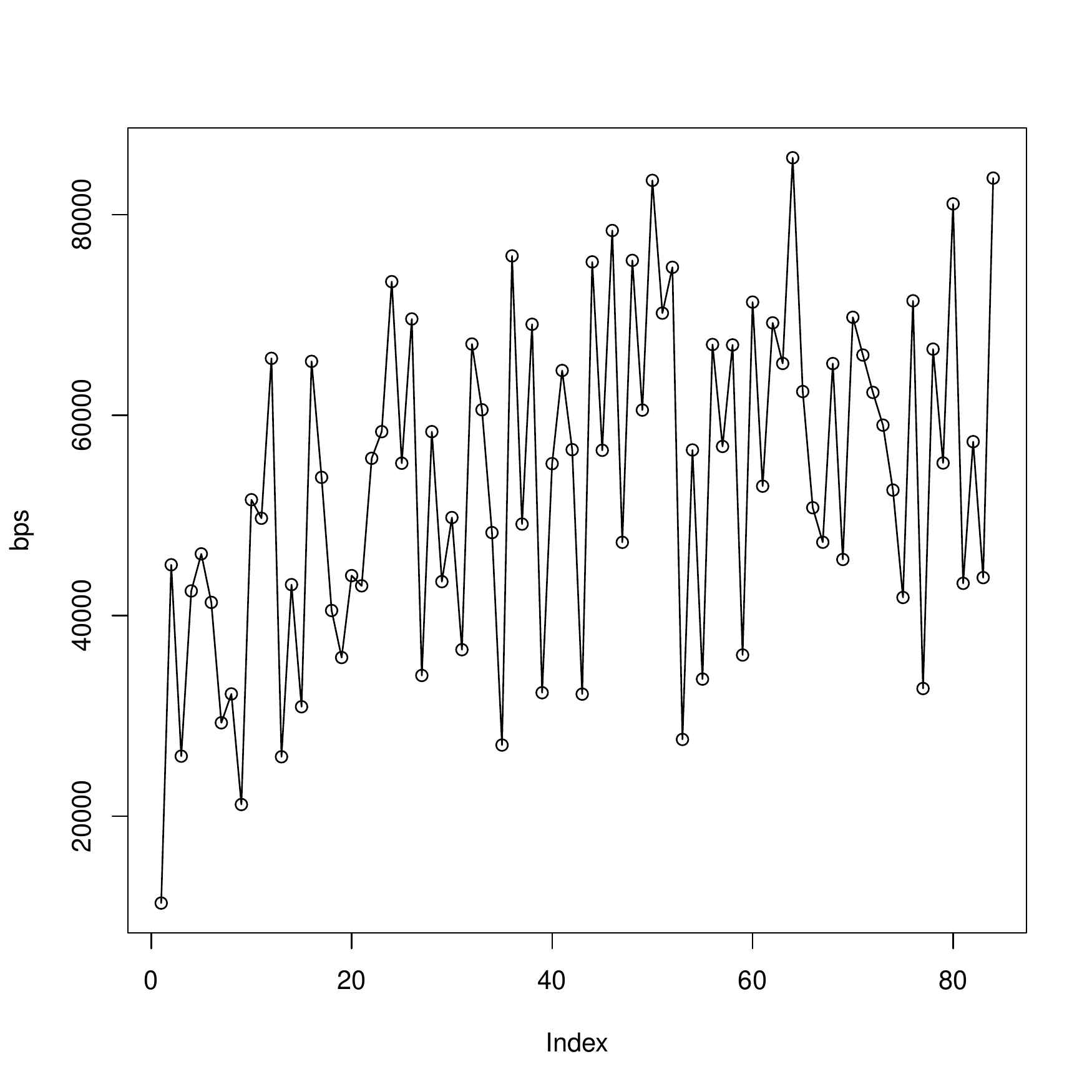}
        \caption{6.51813 Kilobytes (performed with the touchpad).}
        \label{brainbs6nab1}
    \end{subfigure}
        \hskip2em
    \begin{subfigure}{.45\linewidth}
        \centering
        \includegraphics[scale=0.37]{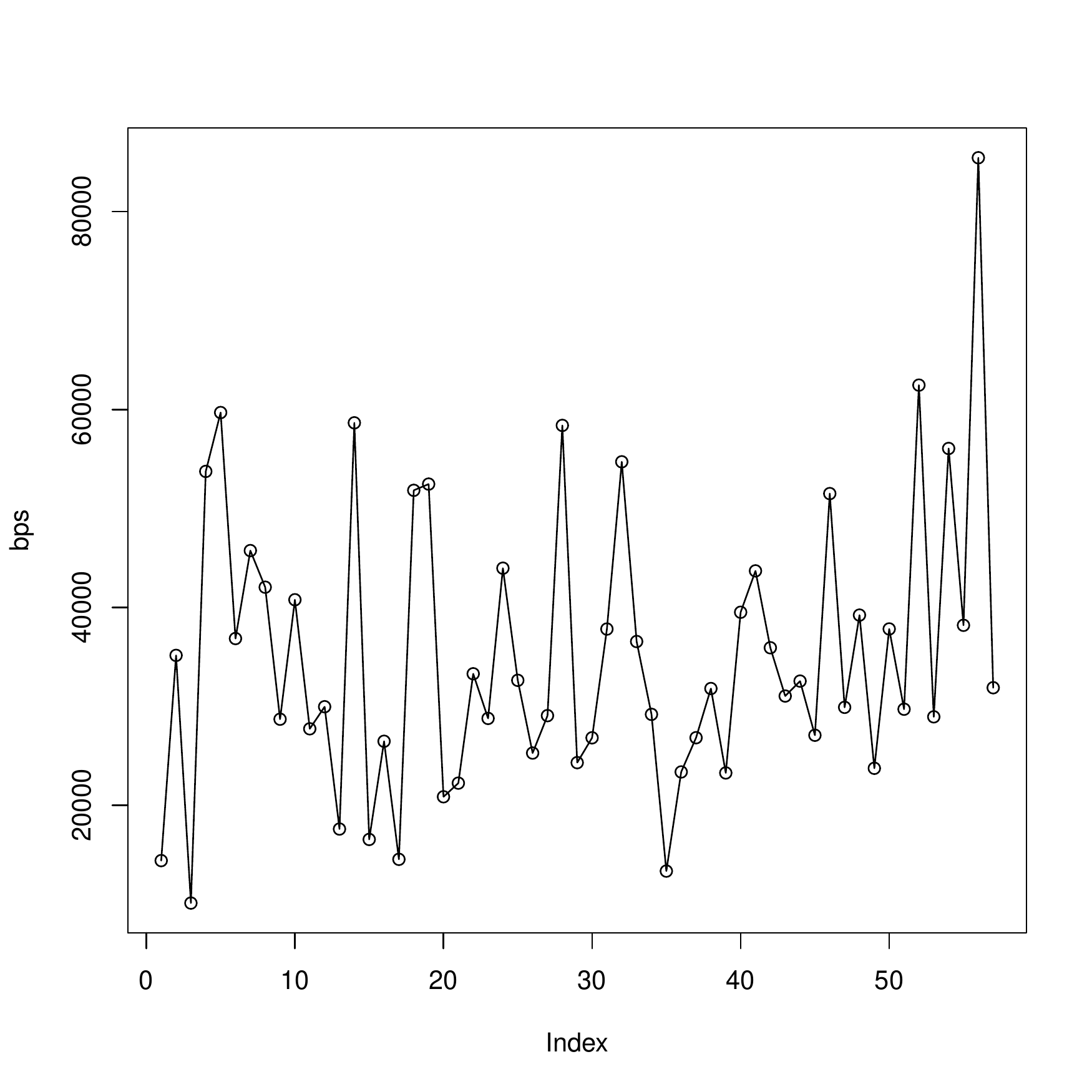}
        \caption{4.31812 Kilobytes (performed with a mouse).}
        \label{brainbs6nab2}
    \end{subfigure}
    \centering
    \begin{subfigure}{.45\linewidth}
        \centering
        \includegraphics[scale=0.37]{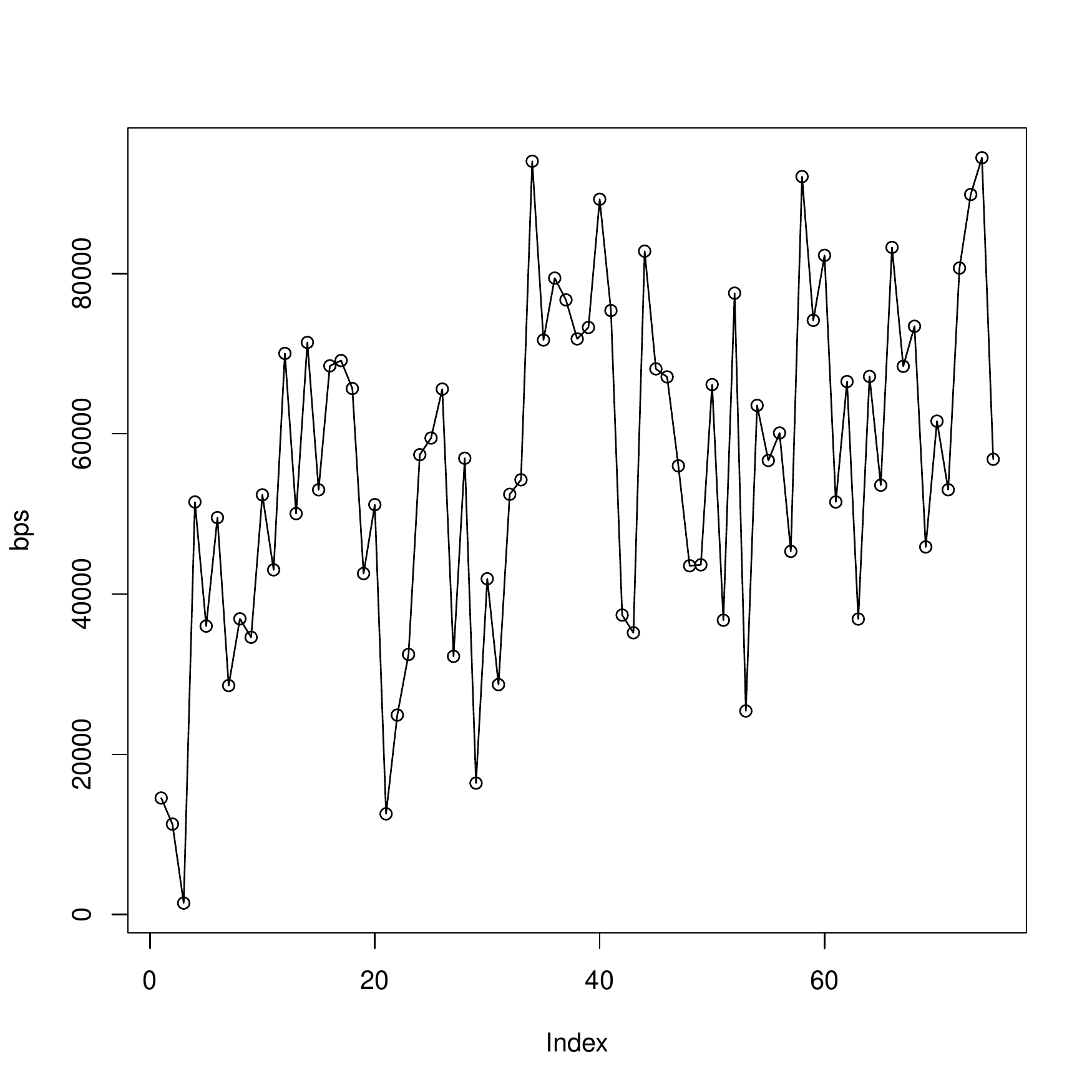}
        \caption{6.79218 Kilobytes (performed with the touchpad).}
        \label{brainbs6nab3}
    \end{subfigure}
    \caption{This test was performed with a male child (10 years old, on the same environment as in Fig  \ref{brainbs6winnb1}). The data and final screenshots can be found at \url{http://smartcity.inf.unideb.hu/~norbi/BrainBSeries6/measurements/NaB/}.  It should be noted that  test subjects with touchpad can use both hands, one for holding the button and the other for motion.}     
    \label{brainbs6nab}
\end{figure*}

\begin{figure*}
    \centering
    \begin{subfigure}{.45\linewidth}
        \centering
        \includegraphics[scale=0.18]{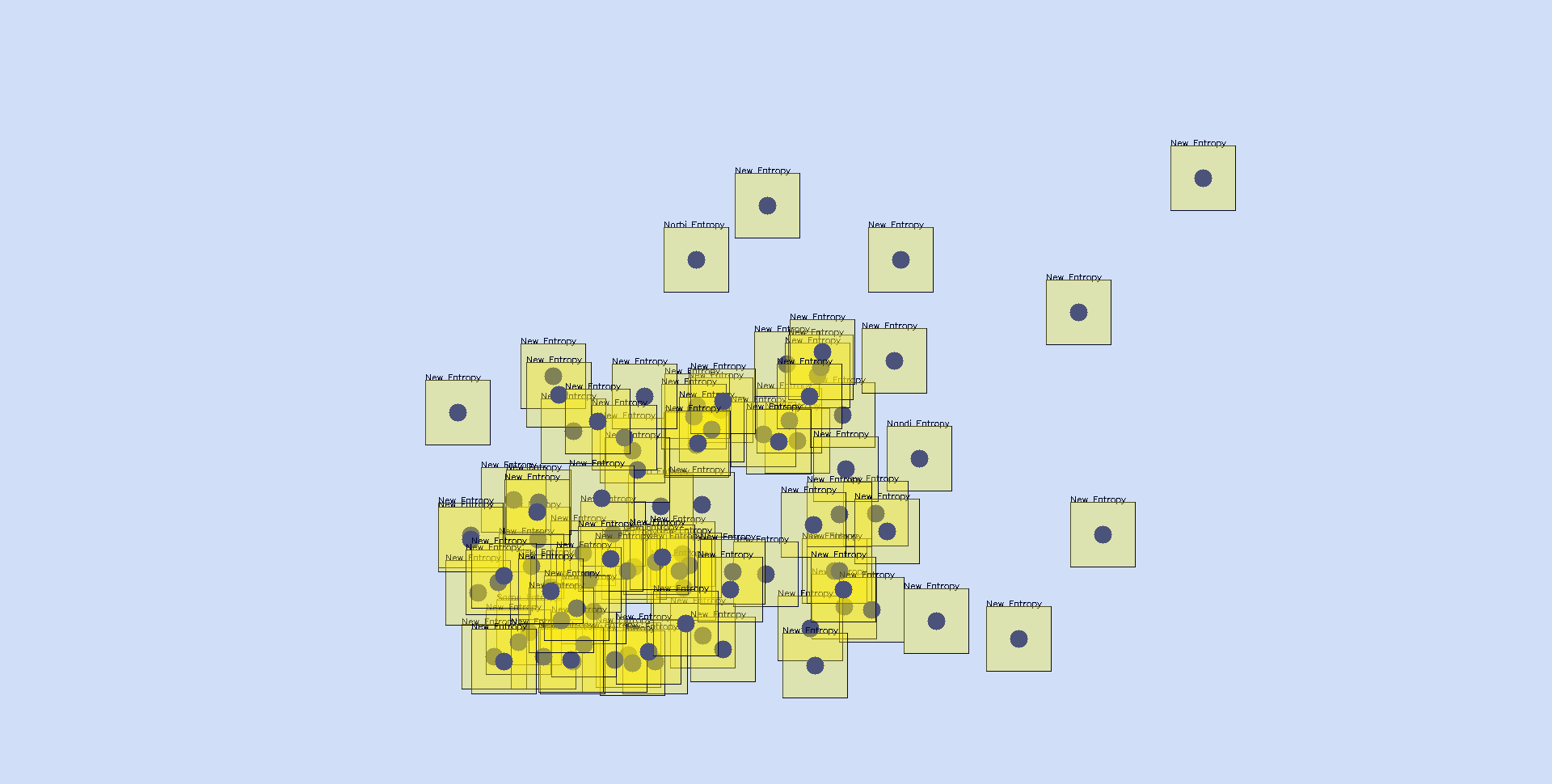}
        \caption{This final screenshot corresponds to Fig \ref{brainbs6nab1}.}
    \end{subfigure}
    \hskip2em
    \centering
    \begin{subfigure}{.45\linewidth}
        \centering
        \includegraphics[scale=0.18]{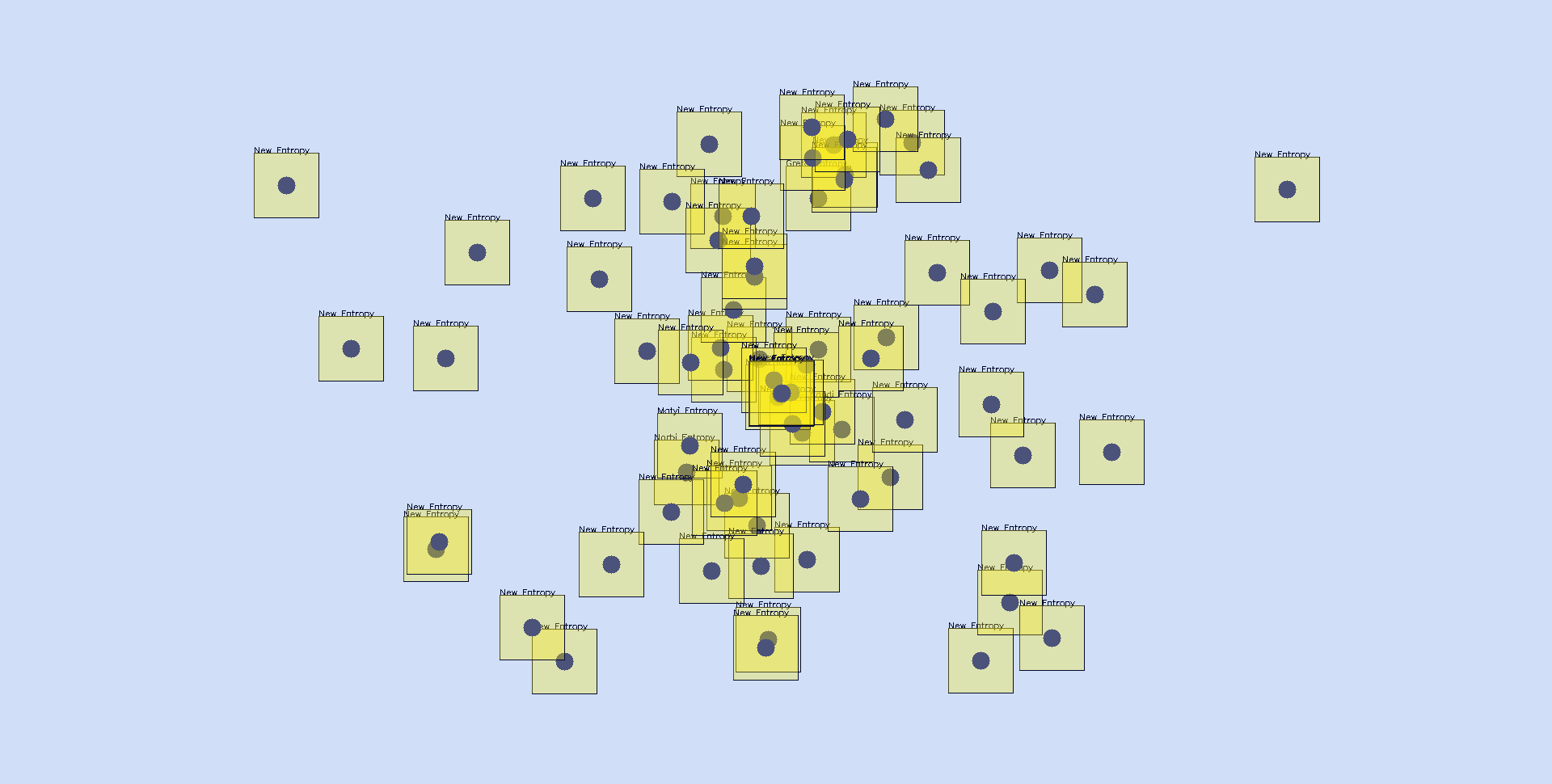}
        \caption{This final screenshot corresponds to Fig \ref{brainbs6nab2}.}
    \end{subfigure}
        \hskip2em
    \begin{subfigure}{.45\linewidth}
        \centering
        \includegraphics[scale=0.18]{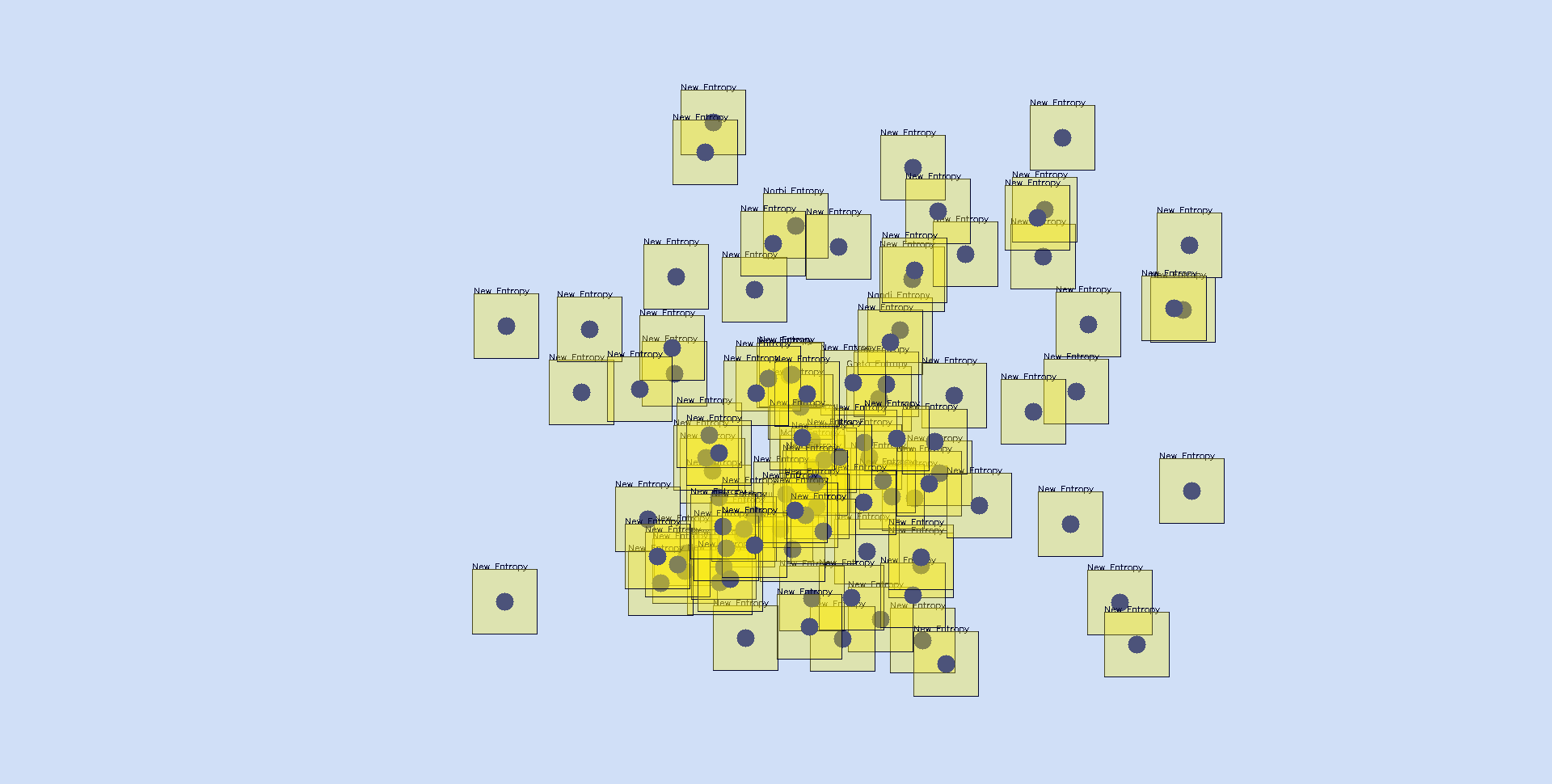}
        \caption{This final screenshot corresponds to Fig \ref{brainbs6nab3}.}
    \end{subfigure}
    \caption{This test was performed with a male child (10 years old, on the same environment as in Fig  \ref{brainbs6winnb1}). The data and final screenshots can be found at \url{http://smartcity.inf.unideb.hu/~norbi/BrainBSeries6/measurements/NaB/}.}
     \label{brainbs6nabs}
\end{figure*}

As concluded in our former preliminary study \cite{brainbs5}, one of the further developments of Series 5 is changing to full screen from fixed-size window. This modification affects the basic operation of the benchmark, 
so the first objective was to verify that whether the feeling of losing the character still appears correctly or not. On Windows systems there were no problems. Some experiments using default settings on Windows 10 can be seen in Fig \ref{brainbs6nb}, Fig \ref{brainbs6nab} and Fig \ref{brainbs6nabs}. But on GNU/Linux systems test subjects reported that the feeling of losing the character is not experienced. These observations will be detailed in a next section.

\subsection{Logging Data}

The state of the BrainB benchmark can be saved at any time by pressing the S button but measured data is automatically saved after the test is finished. The program saves a screenshot of the display to a PNG file. For example, such a screenshot was shown in Fig \ref{brainbs6screen}. The data is saved to a text file that contains the information shown in Listing \ref{savedata}, where the most important lines are the following. 
The two time values tell the time when data is saved. The first one (in Line \ref{time1}) 
is expressed in 100 millisecond units and the second one (in Line \ref{time2}) is expressed in the form minutes:seconds. 
The noc value tells the number of characters (boxes).
The nop value tells the number of pause events initiated by the test subject.
The relation symbol in Line \ref{relation} indicates the fulfillment of our research hypothesis that 
the mean of the complexity of changing from lost to found is less than the mean of the changing found to lost. 

\begin{lstlisting}[language={C},numbers={left},stepnumber={1},numbersep={-1pt},basicstyle={\scriptsize\ttfamily},caption={The structure of the measured data. This log file is belong to the measurement shown in Fig \ref{brainbs6nblog}.},label={savedata},escapechar={|}]
  NEMESPOR BrainB Test 6.0.3
  time      : 6000 |\label{time1}|
  bps       : 28170
  noc       : 71
  nop       : 0
  lost      : 
  30530 31840 39910 10960 60270 71280 50340 51580 31670 
  49260 53710 41620 86830 72580 56310 70560 68870 45500 
  52480 52660 45640 46870 44660 75860 68150 70110 69610 
  47130 61980 75310 90440 75700 62670 54870 69820 75170
  84350 76990 80480 70840 54920 40720 33800 31590 28860 
  24650 27250 53490 58180 56200 57490 53930 39030 83870 
  87180 78270 70990 43600 52360 43910 33820 31120 34830
  32370 32840 37080 32390 
  mean      : 54181
  var       : 18541.5
  found     : 
  12880 22240 26690 11190 19880 36170 14930 28100 25860 
  27580 36040 34590 22250 12060 11760 8880 10660 30840 
  48000 33030 43040 26330 45880 50380 34970 45950 36610 
  46660 47980 45330 65290 57080 55340 54700 43930 34850 
  55030 43240 69500 50770 58680 54750 65470 59610 79030 
  67190 63890 61550 65590 54100 69460 69210 37390 41850 
  53130 31650 45400 46430 50490 44310 35960 53510 25760 
  38950 33250 39360 46650 63050 64890 68590 76430 50570 
  57630 57250 28830 42020 45500 67160 63310 69930 80200 
  76980 56300 44320 58340 79850 81590 69740 88200 89160 
  62640 55030 60510 39810 51660 51730 47720 62330 66150 
  47100 60470 70810 88930 75110 65290 68830 59430 63710 
  22570 36940 29450 43630 53100 55560 64750 39530 59610 
  58250 71950 62800 75250 76720 81910 31730 47010 44890 
  58490 61750 66900 69380 81650 79450 72420 
  mean      : 51442
  var       : 18616.1
  lost2found: 14930 22250 11760 43040 26330 34970 46660 
  43930 50770 61550 54100 37390 31650 44310 25760 50570 
  28830 56300 69740 62640 39810 62330 65290 59430 22570
  39530 31730 72420 
  mean      : 43235
  var       : 16826.7
  found2lost: 31840 10960 60270 51580 31670 49260 53710 
  86830 70560 68870 45500 52660 45640 46870 75860 69610 
  61980 75310 90440 54870 69820 75170 84350 80480 53490 
  56200 83870 78270 
  mean      : 61283
  var       : 18824.2
  mean(lost2found) < mean(found2lost) |\label{relation}|
  time      : 10:0 |\label{time2}|
  U R about 6.37927 Kilobytes
\end{lstlisting}

\subsection{Choosing Colors}

We put a lot of emphasis on what colors to choose for our benchmark. The reason for this is that even the standard test requires a constant focus of 10 minutes, 
which can put a lot of pressure on one’s eyes. To ease this strain as much as we possibly could, we took lots of things into account. Firstly, we tried to maximize the contrast between the background and the figures. This means that we picked some colors that could be easily distinguished and then we ran some manual tests. The result was a significant drop in the overall burden of the eyes.

After this, we thought about how we could make the benchmark available for a wider range of people, namely for those who suffer from 
parachromatism or even
disambiguation. This is rather important as it is said that roughly 8\% of men and 0.5\% of 
women\footnote{\url{http://www.color-blindness.com/2006/04/28/colorblind-population}} suffer from one of these. 
In order for them to be able to comfortably run our benchmark, we tried 
to pick colors that are easily distinguishable even for these people.

Another problem was that we did not target a specific age group. On the contrary, we were especially curious about the results of adults, adolescents, teenagers and children. Therefore, we needed to pick a color scheme that was modern, vivid, yet not too complex and not too abstract. This, too, required a lot of experimentation.

\subsection{Known Problems with Series 6}

\begin{figure*}
    \centering
    \begin{subfigure}{.45\linewidth}
        \centering
        \includegraphics[scale=0.37]{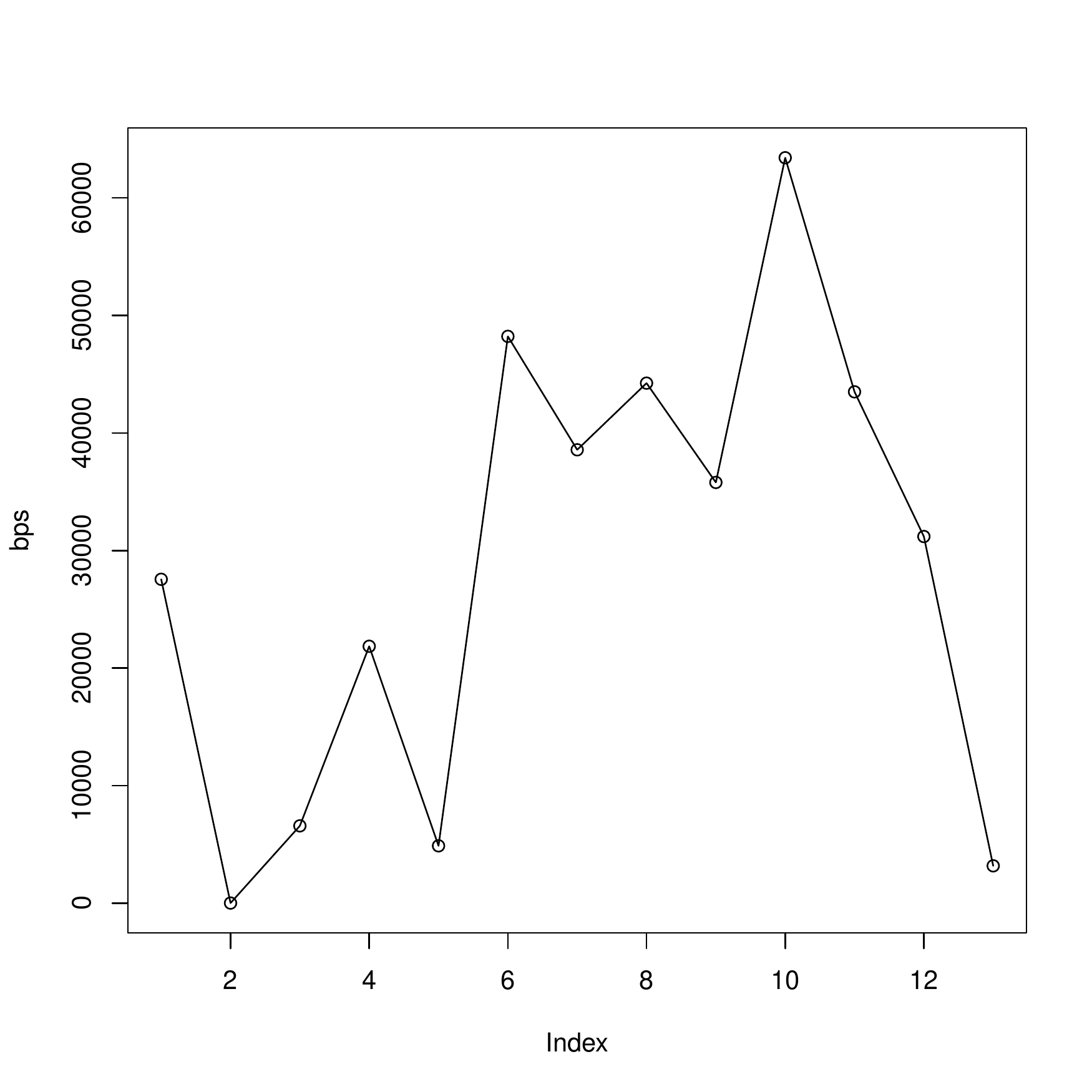}
        \caption{The test subject was the same as in the experiment shown in Fig \ref{brainbs6nb}. The final result was 3.76904 Kilobytes.}
 \label{brainbs6l1}
    \end{subfigure}
        \hskip2em
    \begin{subfigure}{.45\linewidth}
        \centering
        \includegraphics[scale=0.37]{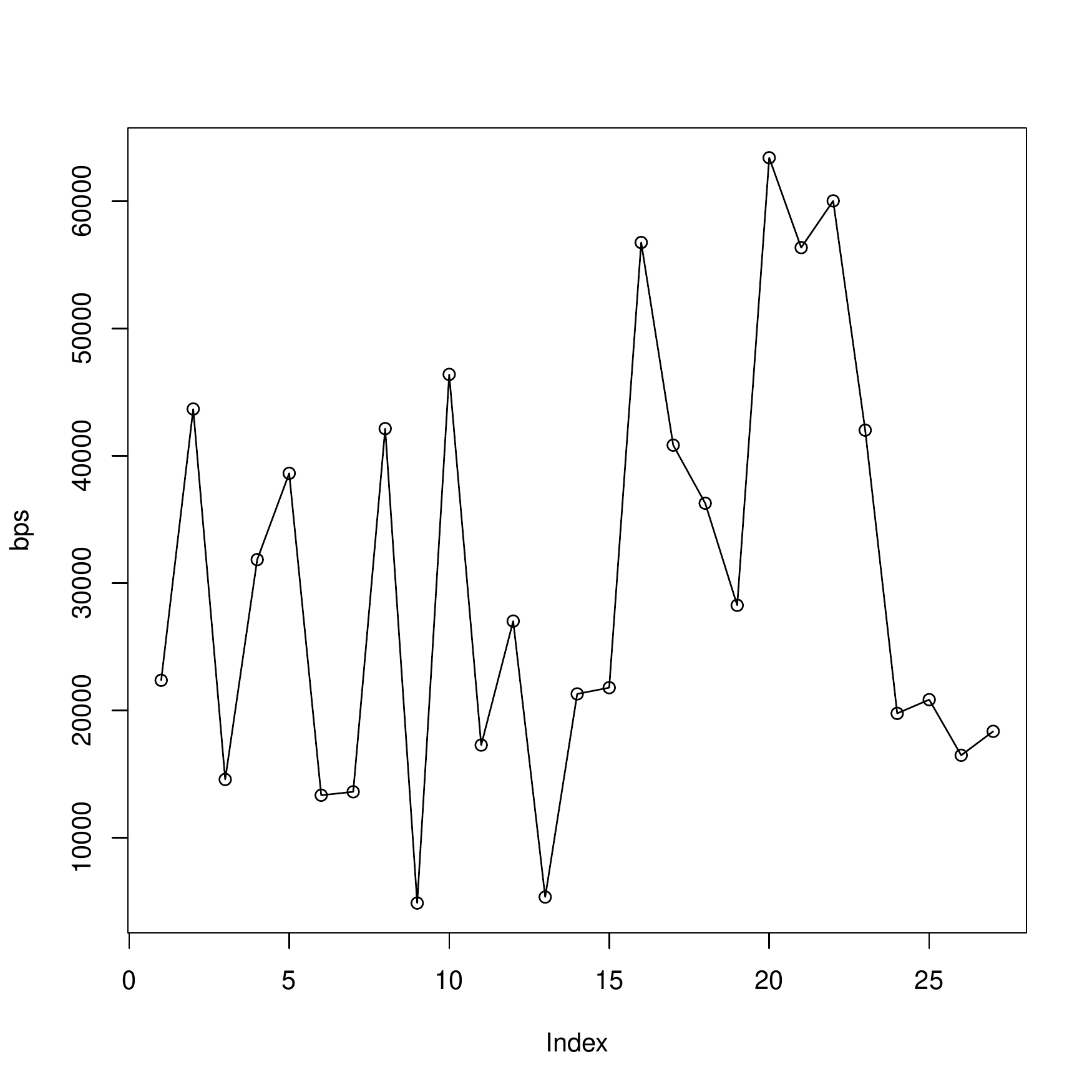}
        \caption{The test subject was the same as in the experiment shown in Fig \ref{brainbs6nab}. The final result was 3.75116 Kilobytes.}
         \label{brainbs6l2}
    \end{subfigure}
    \\
    \centering
    \begin{subfigure}{.45\linewidth}
            \centering
        \includegraphics[scale=0.18]{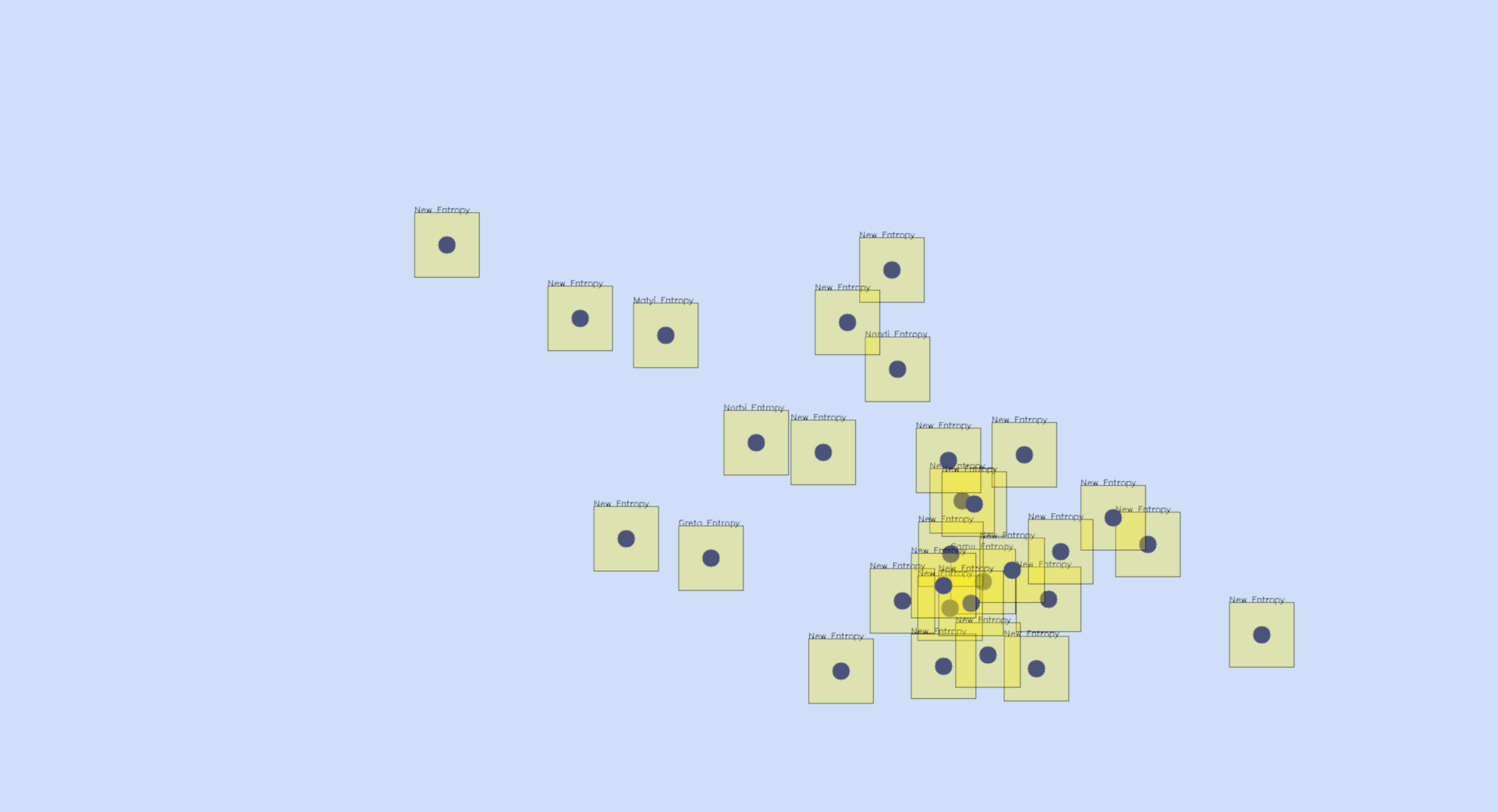}
        \caption{This final screenshot corresponds to Fig \ref{brainbs6l1}.}
         \label{brainbs63}
    \end{subfigure}
    \hskip2em
    \begin{subfigure}{.45\linewidth}
            \centering
        \includegraphics[scale=0.18]{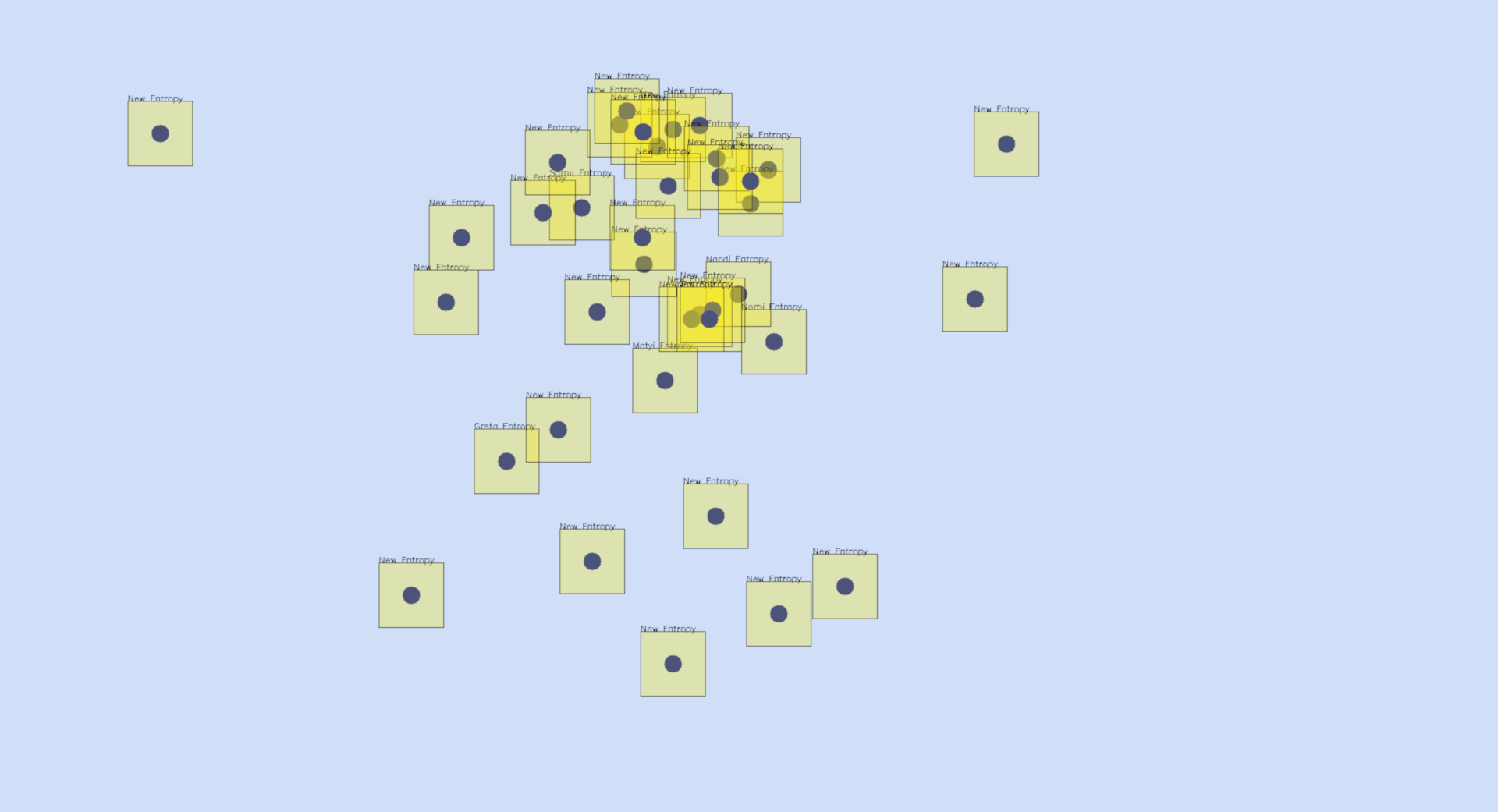}
        \caption{This final screenshot corresponds to Fig \ref{brainbs6l2}.}
         \label{brainbs6l4}
    \end{subfigure}
    \caption{These tests were performed on a GNU/Linux desktop (Ubuntu 16.04, 
    SyncMaster S24B300 monitor with resolution
1920x1080). Test subjects reported that the feeling of losing the character is not experienced.}
    \label{brainbs6l}
\end{figure*}

\begin{figure*}
    \centering
    \begin{subfigure}{.45\linewidth}
        \centering
        \includegraphics[scale=0.37]{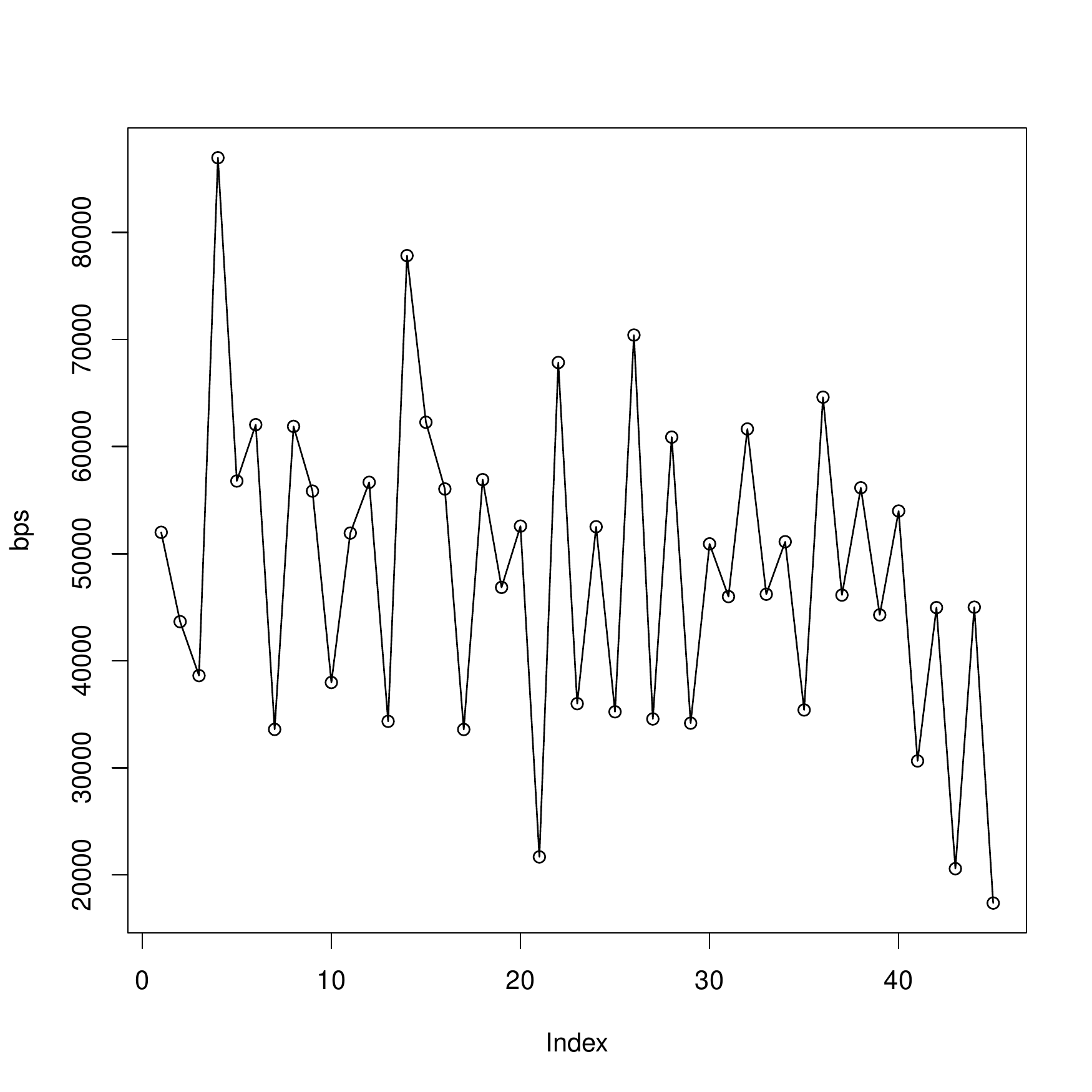}
        \caption{The final result was 5.95587 Kilobytes. xinput settings were the following \enquote{Device Accel Constant Deceleration (277): 1.000000}, \enquote{Device Accel Velocity Scaling (279): 1.000000} and \enquote{Device Accel Profile (276):	-1}}
 \label{brainbs6nbft1}
    \end{subfigure}
        \hskip2em
    \begin{subfigure}{.45\linewidth}
        \centering
        \includegraphics[scale=0.37]{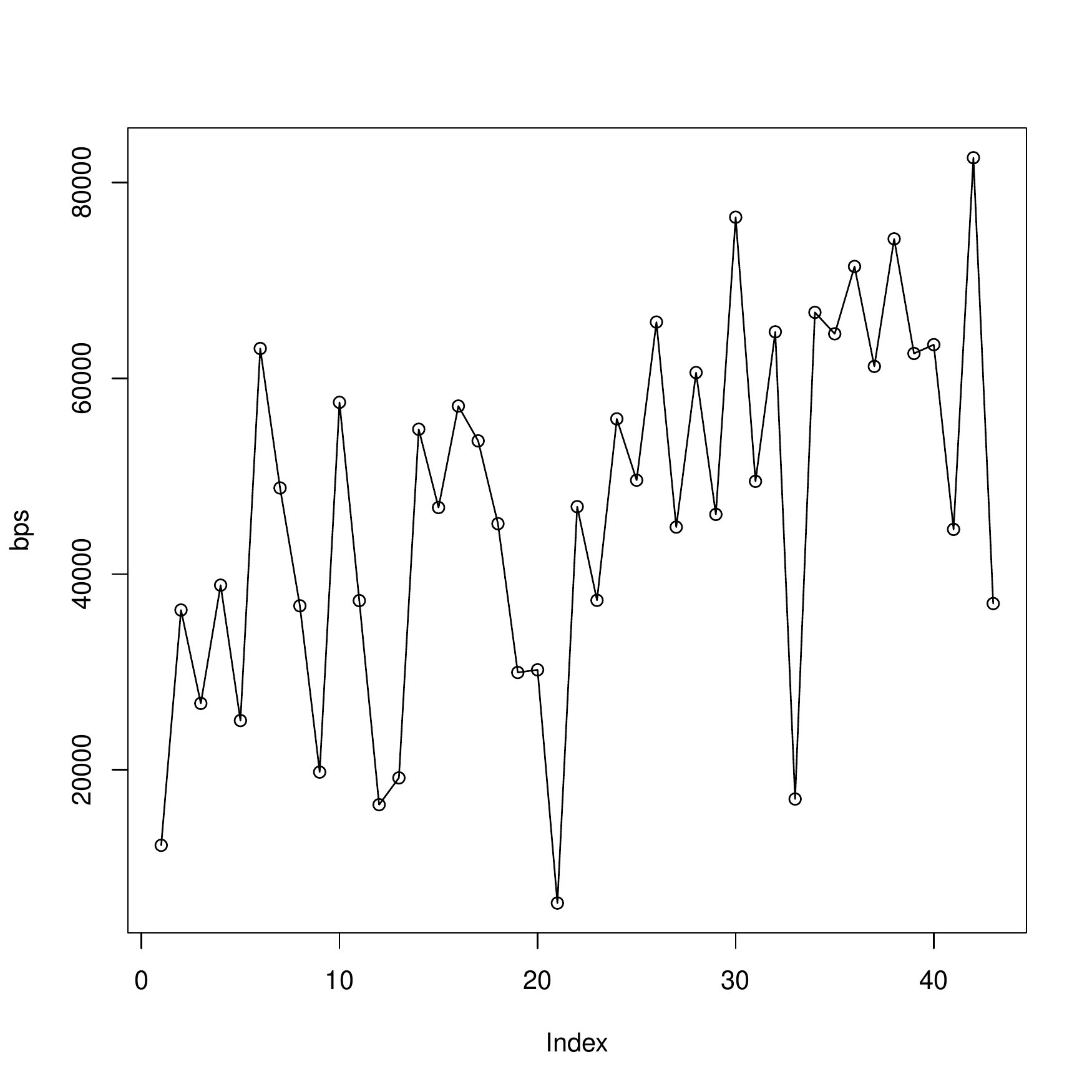}
        \caption{The final result was 5.71674 Kilobytes. xinput settings were the following \enquote{Device Accel Constant Deceleration (277): 2.000000}, \enquote{Device Accel Velocity Scaling (279): 15.000000} and \enquote{Device Accel Profile (276):	-1}}
         \label{brainbs6nbft2}
    \end{subfigure}
    \\
    \centering
    \begin{subfigure}{.45\linewidth}
            \centering
        \includegraphics[scale=0.18]{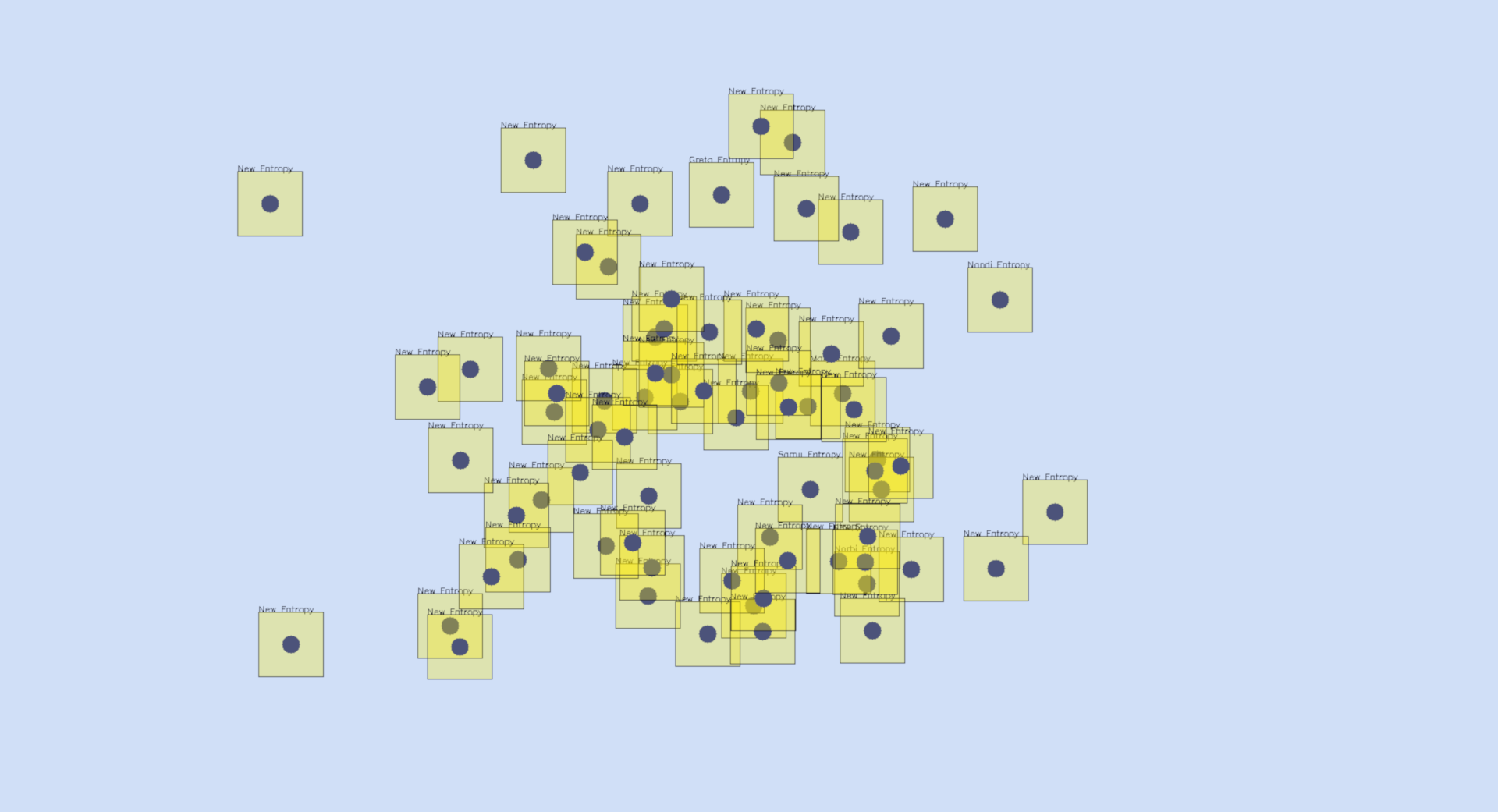}
        \caption{This final screenshot corresponds to Fig \ref{brainbs6nbft1}.}        
         \label{brainbs6nbft3}
    \end{subfigure}
    \hskip2em
    \begin{subfigure}{.45\linewidth}
            \centering
        \includegraphics[scale=0.18]{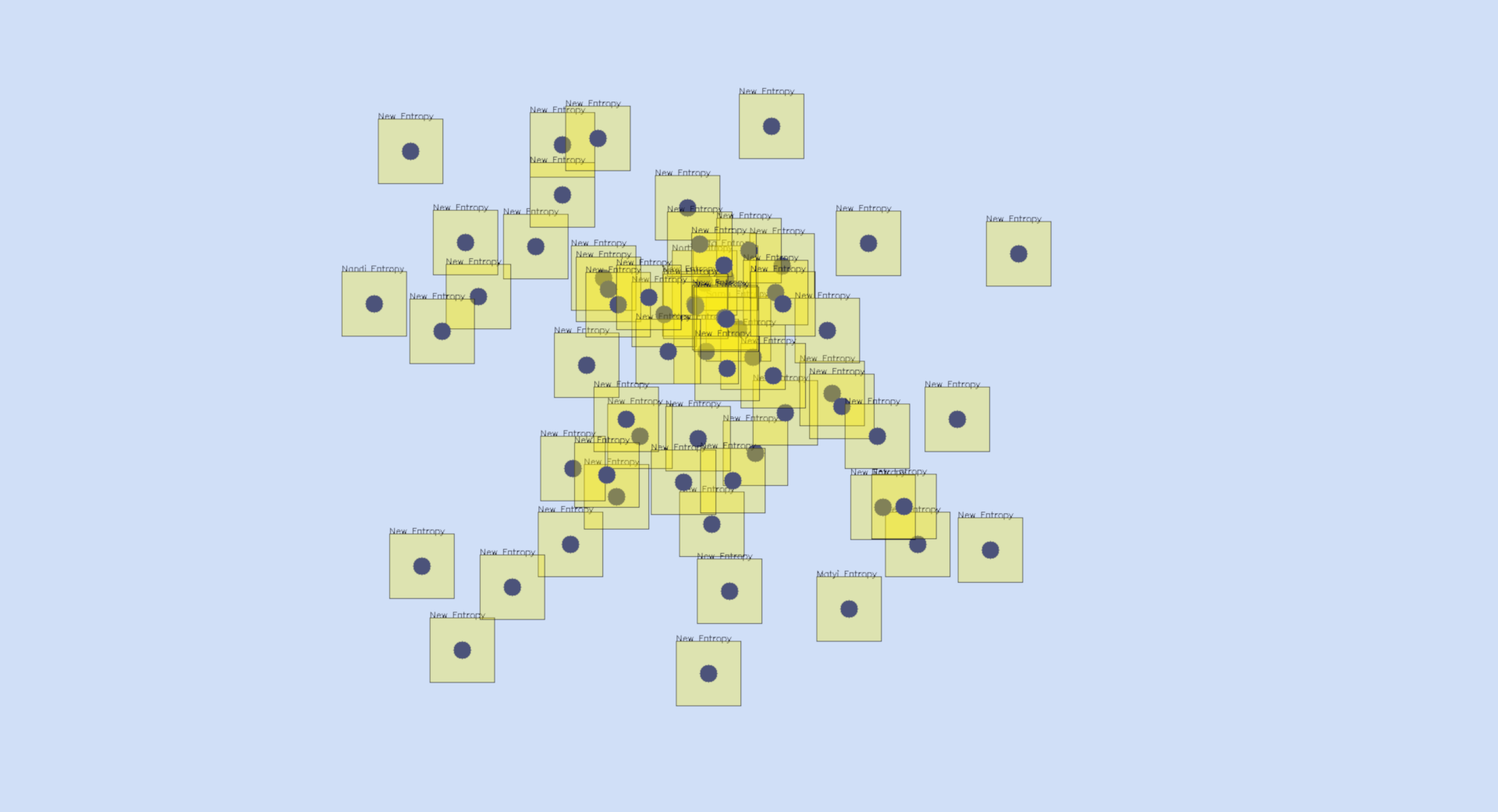}
        \caption{This final screenshot corresponds to Fig \ref{brainbs6nbft2}.}
         \label{brainbs6nbft4}
    \end{subfigure}
    \caption{The test subject was the same as in the experiment shown in Fig \ref{brainbs6nb}. 
    The subject reported that the feeling of losing the character has already been experienced.}    
    \label{brainbs6nbft}
\end{figure*}

Despite that our benchmark is developed on Linux it is surprising that test subjects who performed it on Linux did not experience the feeling of losing the character. This problem causes the deteriorated results shown in Fig \ref{brainbs6l}. It is important to note that it has not been detected in earlier series of the application. Moreover, before Series 6, there was no Windows binary edition of BrainB program. In Series 6, changing to full screen from windowed causes the problem because Series 6  is sensitive to the different mouse sensitivity settings on Windows and Linux systems (the measurements shown in Fig  \ref{brainbs6l} were performed with a Logitech mouse with \textit{acceleration:  5/1} and \textit{threshold:  5} xset m\footnote{\url{https://www.x.org/archive/X11R7.7/doc/man/man1/xset.1.xhtml}} setting). 
A short-term solution may be to standardize the test environment used by each member 
of a given subset of test subjects. We apply this method to perform systematic measurements with Series 6 in the next section. The long-term solution will be to fine-tune the control of movements of boxes that is hardwired into the Series 6 from Series 5 at this moment.
Another possibility is to take the liberty of fine-tuning of the mouse for test subjects who thus would be able to choose their custom mouse settings in order to increase their effectiveness. This is also in well accordance with the competitive way of performing our test.  Fig \ref{brainbs6nbft} presents two measurements using custom mouse settings.

\subsection{Systematic Measurements with Series 6}

The BrainB Series 6 was measured in two groups: UDPROG and DEAC-Hackers.
The first one is a Facebook community of over 550 actual or former students of the BSc course of “High Level Programming Languages” at the University of Debrecen. 
The second one is an esport department of the University of Debrecen's Athletic Club. Participation in the BrainB Series 6 survey was voluntary in both groups. 

\begin{figure*}
    \centering
    \begin{subfigure}{.45\linewidth}
  \centering
    \includegraphics[scale=0.37]{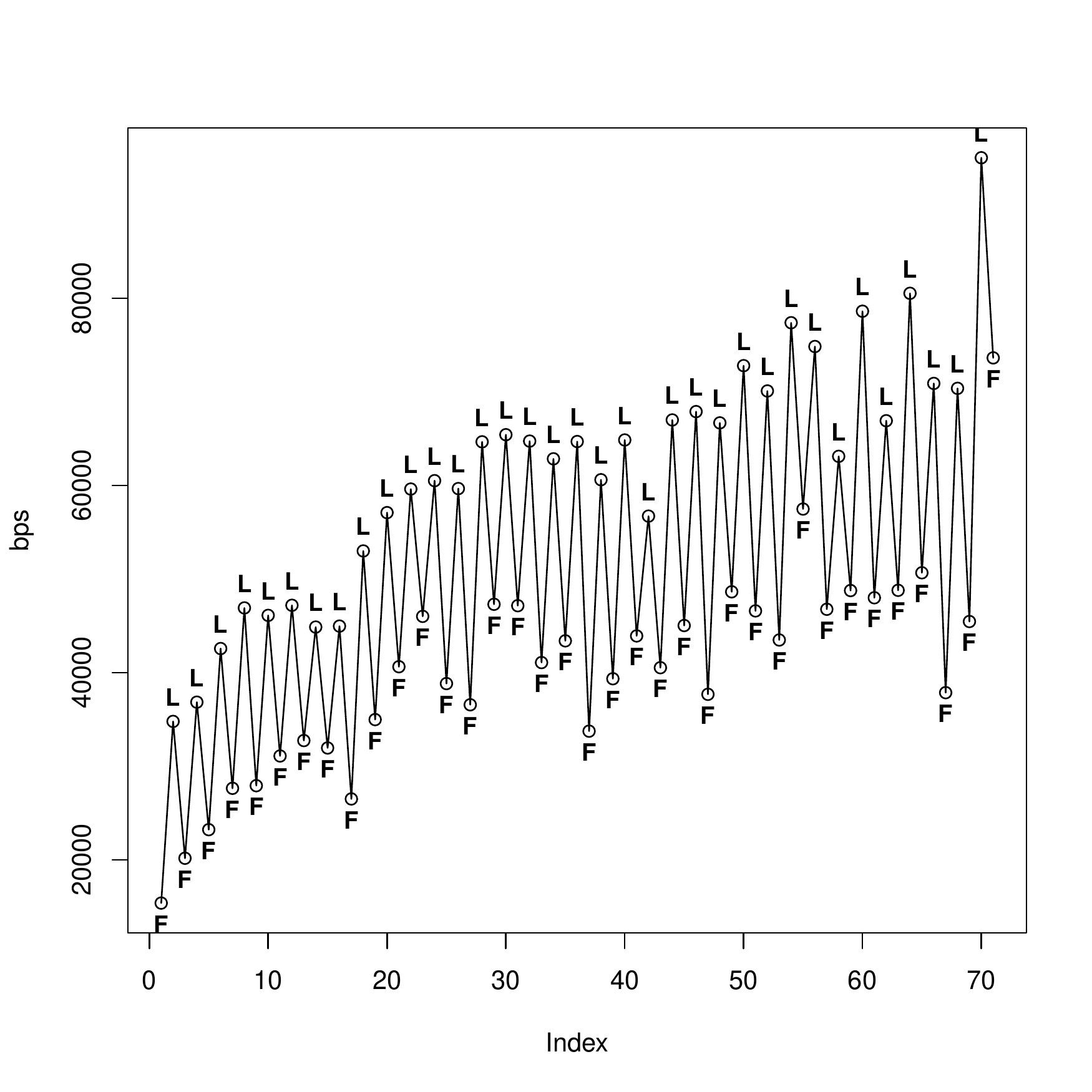}
  \caption{This figure shows the averaged losing and finding curve for all UDPROG participants where the losing (L) and finding (F) events are also  indicated.}
   \label{udprogmeanFL}
    \end{subfigure}
        \hskip2em
    \begin{subfigure}{.45\linewidth}
        \centering
    \includegraphics[scale=0.37]{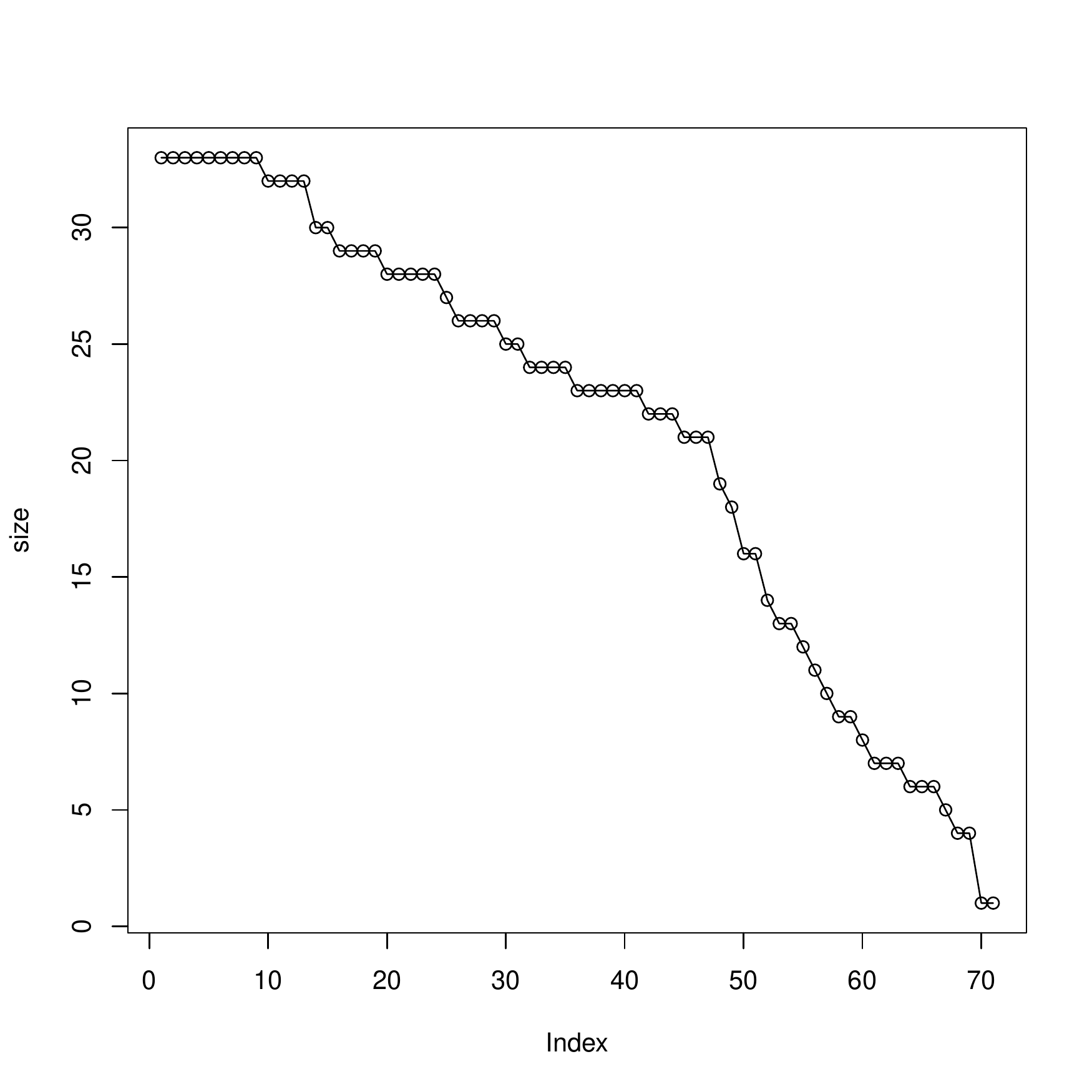}
  \caption{The sizes of samples of losing and finding events. The x-axis shows the sizes and the y-axis shows the number of test-subjects.}
   \label{udprognum}
    \end{subfigure}
    \caption{Measurements in the community UDPROG. The arithmetic mean of the final results of UDPROG participants is 4.95345. 
    The mean of the number of boxes at the moment when the benchmark ends is 57.1818. The anonymized data can be found at \url{http://smartcity.inf.unideb.hu/~norbi/BrainBSeries6/measurements/UDPROG/}.}    
    \label{udprog}
\end{figure*}

In the UDPROG community 33 members send back their results including the PNG screenshot and the produced text file 
within 2 days from the date of announcement (20 August 2018). 
The arithmetic mean of the final results of UDPROG participants is 4.95345. 
The mean of the number of boxes at the moment when the benchmark ends is 57.1818.
The averaged losing and finding curve for all members is shown in Fig \ref{udprogmeanFL}.
At the end of the curve the arithmetic mean values of complexity of the losing and finding events are irrelevant because the size of the sequences of losing and finding events 
are different for every participants. Fig \ref{udprognum} indicates these different sizes.

\begin{figure*}
    \centering
    \begin{subfigure}{.45\linewidth}
  \centering
    \includegraphics[scale=0.37]{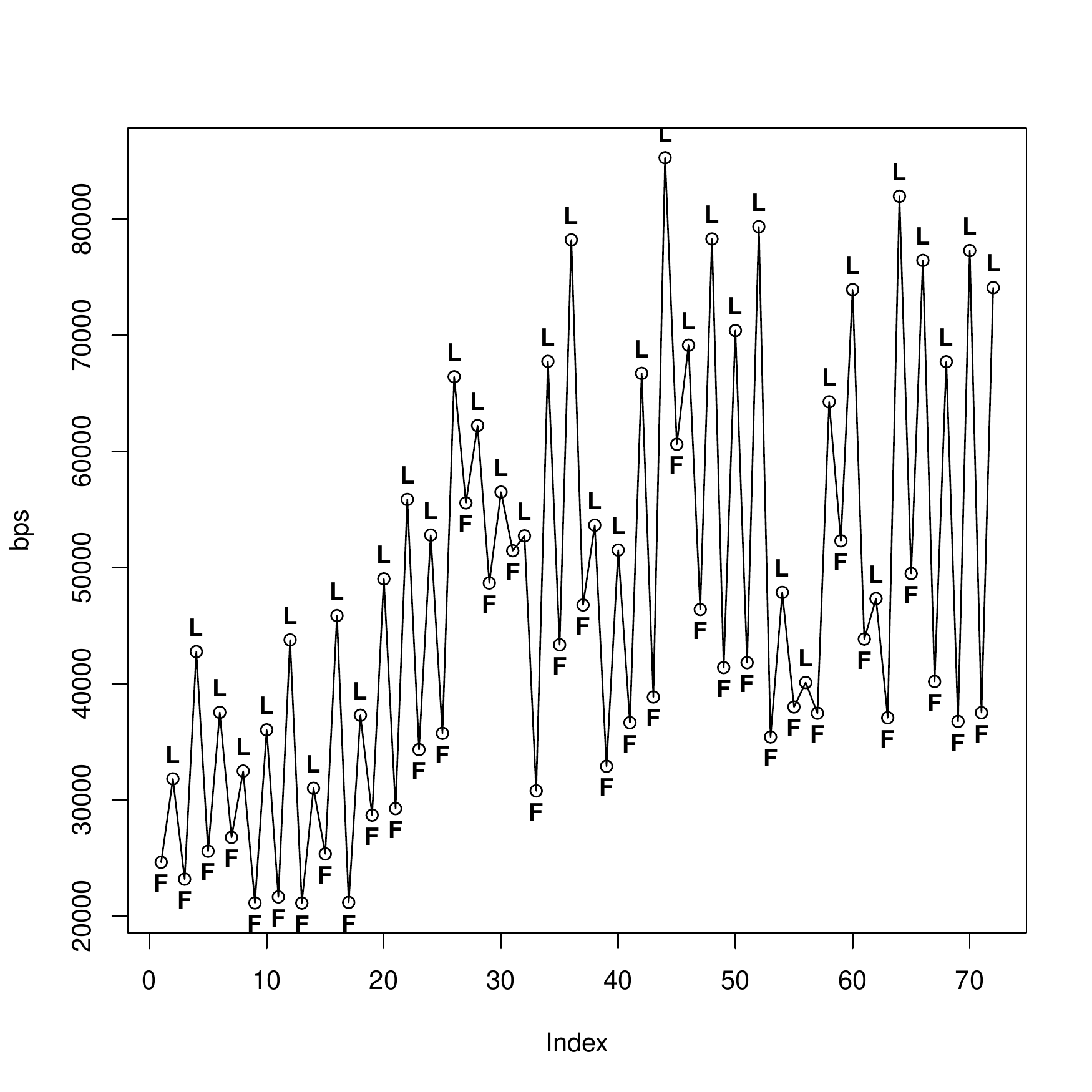}
  \caption{This figure shows the averaged losing and finding curve for all DEAC-Hackers participants where the losing (L) and finding (F) events are also  indicated.}
   \label{deacmeanFL}
    \end{subfigure}
        \hskip2em
    \begin{subfigure}{.45\linewidth}
        \centering
    \includegraphics[scale=0.37]{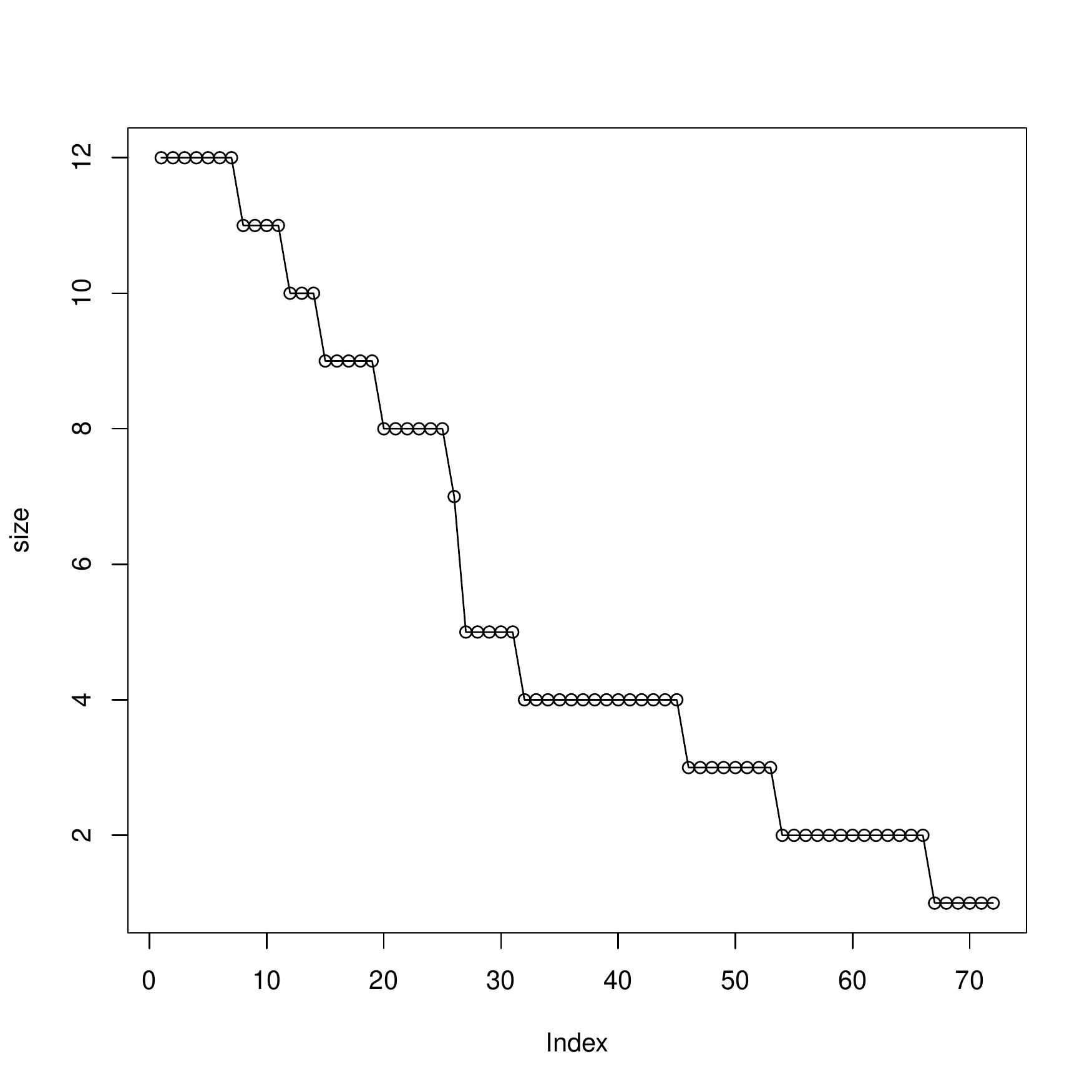}
  \caption{The sizes of samples of losing and finding events. The x-axis shows the sizes and the y-axis shows the number of test-subjects.}
   \label{deacnum}
    \end{subfigure}
    \caption{Measurements in the community DEAC-Hackers. The arithmetic mean of the final results of DEAC-Hackers  participants is 3.71036. It is surprisingly lower than expected if compared to the value 4.95345 of the examined programming community. 
    The mean of the number of boxes at the moment when the benchmark ends is 49. The anonymized data can be found at \url{http://smartcity.inf.unideb.hu/~norbi/BrainBSeries6/measurements/DEACH/}.}    
    \label{deac}
\end{figure*}

In the DEAC-Hackers community 12 esport athletes have sent back their results that can be seen in Fig \ref{deac}. 
It is important to notice that despite low sample sizes of test subjects the averaged losing and finding 
curves shown in Fig \ref{udprogmeanFL} and \ref{deacmeanFL} have already separated the losing and finding events.

\section{Conclusion}

Our research hypothesis was that 
the mean of the complexity of changing lost to found is less than the mean of the changing found to lost. 
Fig \ref{udprogmeanFL} and \ref{deacmeanFL} show the fulfillment of this hypothesis. 
It seems very well in these figures that the averaged losing and finding curve has precisely separated 
the losing and finding events. Intuitively, this result shows that we lose the character on a higher 
complexity level then we find it on a relatively lower level again. 
This simple hypothesis has been proved by the results of this study. 

In order to further strengthen the completion of our benchmark test in a competitive way in the  following versions we are going to offer to test subjects a little more 
liberty of fine-tuning the settings. The fine-tuning of mouse settings was already mentioned earlier. A further possibility is to allow using custom colors.

The next research objective will be to verify the satisfaction of Hick's law. 
To achieve this goal it is simple enough to compare the complexity of finding and losing events with the time differences 
of these. Unfortunately, the actual version of the BrainB benchmark do not record these timestamps. 
The BrainB Series 7 will contain this feature. Our long-term research goal is to further develop our benchmark to a standard psychological test that can be used for talent search in esport.

\section{Acknowledgement}

Thanks to 
the students of the BSc course titled “High Level Programming Languages” at the University of Debrecen, 
to the members of the NEMESPOR mailing lists \url{https://groups.google.com/forum/#!members/nemespor}, 
to the members of the UDPROG Facebook community \url{https://www.facebook.com/groups/udprog}
and to the members of the DEAC-Hackers esport department \url{http://deac.hu/szakosztalyok/esport}
for their interest and for performing the BrainB Test Series 6.
Special thanks 
to Ren\'at\'o Besenczi for pretesting the Windows release of the BrainB program,
to Roland Paszerbovics, Gerg\H o Hajzer and P\'eter Rozsos for their interest and support
and to N\'andor Benj\'amin B\'atfai for performing the BrainB Test Series 6 on several occasions.

Finally, our thanks go to Dr. P\'eter Jeszenszky for reading the manuscript and suggesting improvements.

Author contributions were the following: 
N. B.\footnote{\label{efop363}This work was supported by the construction EFOP-3.6.3-VEKOP-16-2017-00002. The project was co-financed by the Hungarian Government and the European Social Fund.} 
conceived the idea, developed the benchmark program, collected the data from the UDPROG community and analyzed the measurements. 
D. P.\footnote{The work/publication is supported by the GINOP-2.3.2-15-2016-00062 project.} 
wrote the section \enquote{Psychological Background}. 
G. B.\textsuperscript{\ref{efop363}} wrote the section \enquote{Choosing colors}. 
R. B.\textsuperscript{\ref{efop363}} wrote the section \enquote{ Informatics Background} and collected the data from the DEAC-Hackers.
D. V.\textsuperscript{\ref{efop363}} wrote the  section \enquote{Losing the Character}.
All authors edited and reviewed the final version of the manuscript.


\bibliographystyle{alpha}
\bibliography{brainbs6}

\newcommand{\etalchar}[1]{$^{#1}$}
\begin{thebibliography}{AWRG76}

\bibitem[ABR{\etalchar{+}}13]{anguera2013video}
Joaquin~A Anguera, Jacqueline Boccanfuso, James~L Rintoul, Omar Al-Hashimi,
  Farhoud Faraji, Jacqueline Janowich, Eric Kong, Yudy Larraburo, Christine
  Rolle, Eric Johnston, et~al.
\newblock Video game training enhances cognitive control in older adults.
\newblock {\em Nature}, 501(7465):97, 2013.

\bibitem[AWRG76]{WitkinGoodenough}
Herman A.~Witkin and Donald R.~Goodenough.
\newblock Field dependence revisited.
\newblock pages i--85, 1976.

\bibitem[B\'17]{brainbsw}
Norbert B\'atfai.
\newblock esport-talent-search.
\newblock GitHub repository, 2017.
\newblock \url {https://github.com/nbatfai/esport-talent-search} (visited:
  2018-09-30).

\bibitem[BAM{\etalchar{+}}18]{bediou2018meta}
Benoit Bediou, Deanne~M Adams, Richard~E Mayer, Elizabeth Tipton, C~Shawn
  Green, and Daphne Bavelier.
\newblock Meta-analysis of action video game impact on perceptual, attentional,
  and cognitive skills.
\newblock {\em Psychological bulletin}, 144(1):77, 2018.

\bibitem[BBP{\etalchar{+}}18]{brainbs5}
Norbert B\'atfai, Gerg{\H o} Bogacsovics, Roland Paszerbovics, Asztrik Antal,
  Istv\'an Czev\'ar, Viktor Kelemen, and Ren\'at\'o Besenczi.
\newblock {E-sportol\'ok m\'er\'ese} (\textit{Measuring Esport Athletes}).
\newblock {\em Inform\'aci\'os T\'arsadalom}, 18(1):146--155, 2018.
\newblock Original document in Hungarian \url
  {http://real.mtak.hu/79216/1/it_2018_1_10_batfai_et_al.pdf} (visited:
  2018-09-30).

\bibitem[CGR07]{AltPszichobook}
V.~Cs\'epe, M.~Gy{\H o}ri, and A.~Rag\'o.
\newblock {\em \'Altal\'anos pszichol\'ogia 1. - \'Esz\-lel\'es \'es figyelem}.
\newblock Osiris Kiadó, 2007.

\bibitem[GIF{\etalchar{+}}15]{GEYER2015260}
Jason Geyer, Philip Insel, Faraz Farzin, Daniel Sternberg, Joseph~L. Hardy,
  Michael Scanlon, Dan Mungas, Joel Kramer, R.~Scott Mackin, and Michael~W.
  Weiner.
\newblock Evidence for age-associated cognitive decline from internet game
  scores.
\newblock {\em Alzheimer's \& Dementia: Diagnosis, Assessment \& Disease
  Monitoring}, 1(2):260--267, 2015.

\bibitem[HPR{\etalchar{+}}18]{HOMER201850}
Bruce~D. Homer, Jan~L. Plass, Charles Raffaele, Teresa~M. Ober, and Alisha Ali.
\newblock Improving high school students' executive functions through digital
  game play.
\newblock {\em Computers \& Education}, 117:50--58, 2018.

\bibitem[Mac48]{Mackworth}
N.~H. Mackworth.
\newblock The breakdown of vigilance during prolonged visual search.
\newblock {\em Quarterly Journal of Experimental Psychology}, 1(1):6--21, 1948.

\bibitem[MSH{\etalchar{+}}17]{MOISALA2017204}
M.~Moisala, V.~Salmela, L.~Hietajärvi, S.~Carlson, V.~Vuontela, K.~Lonka,
  K.~Hakkarainen, K.~Salmela-Aro, and K.~Alho.
\newblock Gaming is related to enhanced working memory performance and
  task-related cortical activity.
\newblock {\em Brain Research}, 1655:204--215, 2017.

\bibitem[Nie17]{Nieva}
R.~Nieva.
\newblock Facebook's moonshots: Making brains type and skin hear, 2017.
\newblock \url
  {https://www.cnet.com/news/facebook-f8-building-8-moonshot-projects-zuckerberg-regina-dugan/}
  (visited: 2018-07-31).

\bibitem[PHC15a]{pataki2015computer}
B{\'e}la Pataki, Peter Han{\'a}k, and G{\'a}bor Csukly.
\newblock Computer games for older adults beyond entertainment and training:
  Possible tools for early warnings-concept and proof of concept.
\newblock In {\em ICT4AgeingWell}, pages 285--294, 2015.

\bibitem[PHC15b]{10.1007/978-3-319-27695-3_13}
B{\'e}la Pataki, P{\'e}ter Han{\'a}k, and G{\'a}bor Csukly.
\newblock Surpassing entertainment with computer games: Online tools for early
  warnings of mild cognitive impairment.
\newblock In Markus Helfert, Andreas Holzinger, Martina Ziefle, Ana Fred, John
  O'Donoghue, and Carsten R{\"o}cker, editors, {\em Information and
  Communication Technologies for Ageing Well and e-Health}, pages 217--237.
  Springer International Publishing, 2015.

\bibitem[Pow86]{Powers86}
Donald~E. Powers.
\newblock Test preparation for the gre analytical ability measure: Differential
  effects for subgroups of gre test takers.
\newblock 1986.

\bibitem[PS84]{PowersSwinton}
D.~E. Powers and S.~S. Swinton.
\newblock Effects of self-study for coachable test item types.
\newblock {\em Journal of Educational Psychology}, 76(2):266--278, 1984.

\bibitem[RSA06]{PszichoMeresbook}
Nagyb\'anyai N.~O. R\'ozsa~S. and Ol\'ah A.
\newblock {\em A pszichol\'ogiai m\'er\'es alapjai: Elm\'elet, m\'odszer és
  gyakorlati alkalmaz\'as}.
\newblock B\"olcs\'esz Konzorcium, 2006.

\bibitem[Sag12]{Sagan}
Carl Sagan.
\newblock {\em Dragons of Eden: Speculations on the Evolution of Human
  Intelligence}.
\newblock Random House Publishing Group, 2012.

\bibitem[Seo05]{seow}
Steven~C. Seow.
\newblock Information theoretic models of hci: A comparison of the hick-hyman
  law and fitts' law.
\newblock {\em Human–Computer Interaction}, 20(3):315--352, 2005.

\bibitem[TP13]{ToulousePieron}
E.~Toulouse and H.~Pi\'eron.
\newblock {\em Toulouse-Pi\'eron-Revisado Prueba perceptiva y de atenci\'on}.
\newblock TEA, 8 edition, 2013.
\newblock \url
  {http://www.web.teaediciones.com/Ejemplos/Extracto_libro_TP-R.pdf} (visited:
  2018-07-30).

\bibitem[WLH{\etalchar{+}}54]{WitkinEtAl}
H.~A. Witkin, H.~B. Lewis, M.~Hertzman, K.~Machover, P.~Meissner, and
  S.~Bretnall~Wapner.
\newblock {\em Personality through perception: an experimental and clinical
  study}.
\newblock Harper, Oxford, England, 1954.

\bibitem[YD08]{YerkesDodsonLaw}
Robert~M. Yerkes and John~D. Dodson.
\newblock The relation of strength of stimulus to rapidity of habit-formation.
\newblock {\em Journal of Comparative Neurology and Psychology}, 18:459--482,
  1908.

\end{thebibliography}

\end{document}